# Dielectric, magnetic and lattice dynamics properties of double perovskite $(Ca_{0.5}Mn_{1.5})MnWO_6$

*Hong Dang Nguyen[1,2,*], Alexei A. Belik[3], Petr Kužel[1], Fedir Borodavka[1], Maxim Savinov[1], Klára Beranová[4], Jan Drahokoupil[1], M. Jarošová[1], Petr Proschek[5], Bartoloměj Vaníček[1], Stanislav Kamba[1,*]*

[1]*Institute of Physics, Czech Academy of Sciences, Na Slovance 2, 182 00 Prague 8, Czech Republic*

[2]*Faculty of Nuclear Sciences and Physical Engineering, Czech Technical University in Prague, Břehová 7, 115 19 Prague 1, Czech Republic*

[3]*Research Center for Materials Nanoarchitectonics (MANA), National Institute for Materials Science (NIMS), Namiki 1-1, Tsukuba, Ibaraki 305-0044, Japan*

[4]*Institute of Physics, Czech Academy of Sciences, Cukrovarnická 10, 162 00 Prague 6, Czech Republic*

[5]*Department of Condensed Matter Physics, Faculty of Mathematics and Physics, Charles University, Ke Karlovu 5, Prague 2, 121 16 Czech Republic*

**Corresponding author:**

Stanislav Kamba
Institute of Physics,
Czech Academy of Sciences,
Na Slovance 2,
182 00 Prague 8,
Czech Republic
kamba@fzu.cz

## ABSTRACT

Recent dielectric and magnetic studies of $(Ca_{0.5}Mn_{1.5})MnWO_6$ ceramics [A.A. Belik, *Chem. Mater.* **36**, 7604 (2024)] have classified this material as a rare hybrid multiferroic, with both antiferromagnetic and (anti)ferroelectric ordering occurring at the same temperature of 22 K. The pronounced dielectric anomaly observed at this temperature indicated that the structural change is primarily induced by a phonon soft mode and not by a spin arrangement, as is usually the case in type II multiferroics. However, our comprehensive investigation involving new ceramic samples as well as the sample from the above-mentioned reference does not support this conclusion. Low-temperature polarization measurements revealed no evidence of either ferroelectric or antiferroelectric order in both sample series. The dielectric permittivity exhibits only a slight change at the antiferromagnetic transition, and phonon modes observed in IR and Raman spectra show no indication of a symmetry change at low temperatures. In the new samples the Néel temperature is shifted to $T_N$ = 18 K. XRD, SEM, EDS and WDS analyses confirmed the composition $(Ca_{0.5}Mn_{1.5})MnWO_6$ of both ceramics, but also indicated a small amount (percentage points) of MnO and CaO impurities in the sample from the previous publication and $Mn_3O_4$, $CaWO_4$ secondary phases (<4%) in the new ceramics. The differences in dielectric and magnetic properties of the two samples can therefore be explained by their different chemical purity. The small dielectric anomaly of the new sample at the antiferromagnetic transition temperature is explained by a spin-phonon coupling. We conclude that $(Ca_{0.5}Mn_{1.5})MnWO_6$ is not a multiferroic, but a paraelectric antiferromagnet.

## 1. Introduction

Magnetoelectric multiferroics, where magnetic and (anti)ferroelectric orders coexist, have garnered massive interest in recent two decades because of their potential usage in multifunctional devices such as sensors, memory storage, and spintronic devices. In particular, multiferroic perovskite oxides $ABO_3$ attracted a significant attention [1, 2] due to their highly tunable physical properties, namely a magnetoelectric coupling, which stem from their versatile chemical and structural characteristics and promise applications in electronic or spintronic devices [3, 4], fuel cells [5], solar cells [6], etc.. Among them, the B-site ordered double-perovskite oxides, generally represented by the formula $A_2BB'O_6$ (where A is a divalent or trivalent metal, B and B' are transition metal ions arranged alternately in a rock-salt structure and surrounded by corner-sharing oxygen octahedra) have awakened intensive research in the past few years [7, 8]. The incorporation of various transition metals into the B-site of perovskite structures has led to the discovery of multiferroic materials that exhibit a strong magnetoelectric coupling [9-11].

$(Ca_{0.5}Mn_{1.5})MnWO_6$ seems to be a promising compound of double perovskite oxide family. It crystallizes in the space group $P2_1/n$ [12] and exhibits an antiferromagnetic (AFM) ordering below $T_N$ = 22 K. The incorporation of Mn both at the A- and B-perovskite sites creates a complex interplay between the magnetic and lattice subsystems, leading to possible spin-lattice interactions and lattice distortion. Indeed, the dielectric permittivity $\varepsilon'(T)$ exhibits a Curie-Weiss-like growth upon cooling down to 22 K and below a sharp drop was observed similar to the dielectric anomaly at (anti)ferroelectric phase transitions [12]. Absence of dielectric loss maximum at 22 K together with a hypothetical negative Curie-Weiss temperature $\Theta$ = - 128 K obtained from the fit of $\varepsilon'(T)$, could indicate the coexistence of antiferroelectric and AFM phase transitions at the same temperature. In type II multiferroics, the critical magnetic and ferroelectric temperatures may be identical, but in this case a weak ferroelectric polarization is induced by an interaction between spins which appears to be the primary driving force of both phase transitions. However, in the case of $(Ca_{0.5}Mn_{1.5})MnWO_6$, the observed Curie-Weiss behavior of the permittivity may indicate a displacive structural phase transition playing the role of a leading mechanism. The AFM arrangement then could arise due to this change in the crystal structure. In such a case, $(Ca_{0.5}Mn_{1.5})MnWO_6$ would belong to the rare type III hybrid multiferroic systems, similar to the recently reported quadruple perovskite $BiMn_3Cr_4O_{12}$ [13], where the structural change triggers the magnetic order.

In this study, we systematically re-investigate dielectric, magnetic, and structural properties of new $(Ca_{0.5}Mn_{1.5})MnWO_6$ ceramic samples prepared using the same method as in ref. [12]. In addition, we study lattice dynamics using infrared, Raman, and THz spectroscopy for the first time and searching for ferroelectric polarization when measuring pyroelectric current and ferroelectric hysteresis loops. Our findings reveal that $(Ca_{0.5}Mn_{1.5})MnWO_6$ exhibits AFM phase transition with $T_N$ = 18 K, i.e. 4 K lower than in ref. [12]. However, contrary to ref. [12], no signature of ferroelectric or antiferroelectric order is observed. Indeed, phonons in the infrared (IR) and Raman spectra do not indicate any structural change, the temperature behavior of the permittivity is not found to obey the Curie-Weiss behavior, and no spontaneous polarization is detected below $T_N$. A small dielectric anomaly observed at $T_N$ is explained by a spin-phonon coupling.

We also carefully analyzed the chemical composition of the original sample from ref. [15] and compared it with that of the new samples. All samples show some degree of contamination with secondary phases, which are most likely responsible for the different $T_N$ values and different dielectric properties.

## 2. Experiments

Bulk polycrystalline ceramic $(Ca_{0.5}Mn_{1.5})MnWO_6$ samples were synthesized from stoichiometric mixture of $CaWO_4$ and MnO (99.9%). The sample preparation was similar to that described in ref [12]. The synthesis was operated at roughly 6 GPa and 1550 K for 2h in Au capsules utilizing a belt-type high-pressure instrument. After annealing at 1550 K, the samples were rapidly cooled down to room temperature as the heating current was turned off and the pressure gradually released. Hard pellets of about 5 mm in diameter were obtained after opening the capsules. For dielectric, THz, IR, and Raman studies the ceramic discs were polished on both sides, for magnetic measurements using SQUID, the samples were cut to dimensions of 3.6x3.6x0.5 $mm^3$.

The structure and chemical composition of our ceramics was verified by X-ray powder diffraction (XRPD) and wave dispersion spectroscopy (WDS) (see Fig. S2). The chemical composition was measured by the Electron Probe microanalyzer JEOL JXA-8230. The device is equipped with 5 wavelength-dispersive spectrometers and an energy-dispersive spectrometer (EDS) Bruker QUANTAX 200. Additionally, detectors of secondary electrons and back-scattered electrons (BSE) for imaging are induced. For the energy dispersive spectroscopy (EDS) analyses, a standardless Phi-Rho-Z quantification model was used, and oxygen was calculated for stoichiometric composition. WDS was performed using the ZAF correction method, employing $Y_3Fe_5O_{12}$, $CaMg(SiO_3)_2$, and pure W and Mn as standards. The X-ray diffraction experiment was performed using an X'Pert diffractometer (Empyrian) in Bragg-Brentano geometry, with Cu radiation (λ = 1.54056 Å), a 0.5° divergence slit, and a linear detector.

Unpolarized Raman measurements were performed via Renishaw RM1000 Micro-Raman spectrometer equipped with Bragg filters and an $Ar^+$ ion laser with a wavelength of 514.5 nm. The measurements were taken in a backscattering configuration over the wave number range of 5-1800 $cm^{-1}$ and the temperature-dependent Raman spectra were obtained over a temperature range of 5 K to 300 K using an Oxford Instruments Microstat continuous-flow optical He cryostat.

Dielectric properties were measured in a broad range of frequencies from 1 Hz to 950 kHz using Novocontrol Alpha-AN high-performance impedance analyzer in conjunction with a Janis ST-100 cryostat (8 - 300 K). The temperature rate and the alternating electric field were about 3 $Kmin^{-1}$ and 1 $Vcm^{-1}$, respectively. The experimental specimen was prepared as a 400 μm thick, plane-parallel polished plate. Contacts for the electric field application were established utilizing silver wires affixed to the electrodes with silver paste. Thermally stimulated depolarization current measurements were conducted with a KEITHLEY 617 Electrometer. The sample was first cooled down to 8 K under a poling electric field, then the electric field was switched off and the pyroelectric current was measured during the heating. This was repeated for several values of the poling field. At selected temperatures, we also attempted to measure ferroelectric hysteresis loops with a field up to 10 kV/cm. The P–E loops were measured at a frequency of 50 Hz using home-made Sowyer-Tower bridge.

The complex transmittance in the terahertz (THz) range was assessed utilizing a custom-built time-domain spectrometer, which is driven by a Ti:sapphire femtosecond laser emitting 35-fs pulses centered at 800 nm. A photoconductive switch was used as an emitter and an electro-optic sampling scheme using 1-mm-thick, (110)-oriented ZnTe was employed for the phase-sensitive detection of the THz pulses. This approach enables a direct calculation of the complex refractive index and dielectric permittivity spectra in the THz range. [14]

Low-temperature IR reflectivity measurements were conducted using a Bruker IFS-113v Fourier-transform IR spectrometer, which is equipped with a liquid-helium-cooled Si bolometer operating at 1.6 K as the detector.

The temperature control for both the THz complex transmittance and IR reflectivity experiments was achieved via Oxford Instruments Optistat optical continuous helium-flow cryostats, featuring mylar (THz) or polyethylene (IR) windows. To perform a common fit of the IR and THz spectra, we used a damped oscillator model to describe the complex permittivity: [15]

$$\tilde{\varepsilon}(\omega) = \varepsilon'(\omega) - i\varepsilon''(\omega) = \varepsilon_\infty + \sum_{j=1}^{n} \frac{\Delta\varepsilon_j \omega_{TOj}^2}{\omega_{TOj}^2 - \omega^2 + i\omega\gamma_j}, \qquad (1)$$

where $\Delta\varepsilon_j$, $\omega_{TOj}$ and $\gamma_j$ stand for the dielectric strength, frequency and damping of the $j$-th polar phonon, respectively; $\varepsilon_\infty$ denotes the high-frequency (electronic) permittivity it was determined from the room temperature frequency-independent reflectivity tail in the middle IR region, and it was considered to be temperature independent.

The reflectivity, $R(\omega)$, is related to the complex permittivity, $\tilde{\varepsilon}(\omega)$, via:

$$R(\omega) = \left| \frac{\sqrt{\tilde{\varepsilon}(\omega)} - 1}{\sqrt{\tilde{\varepsilon}(\omega)} + 1} \right|^2 \qquad (2)$$

The magnetic properties of the sample were investigated utilizing a vibrating sample magnetometer (VSM, Quantum Design) in conjunction with a Quantum Design physical properties measurement system (PPMS), down to 2 K and using magnetic fields up to 9 T.

The chemical composition of the sample was characterized by X-ray photoelectron spectroscopy (XPS) using the AXIS Supra photoelectron spectrometer (Kratos Analytical). To expose the material unaffected by prior treatment and surface contamination, the portion of the sample was crushed between filter cellulose papers to a fine powder. The powder was spread over a piece of an adhesive paper tape so that the sample formed a thin continuous layer insulated from the sample holder. The XPS measurements were performed using the monochromatized Al Kα X-ray source with 15 mA emission current and the 54° incident beam angle (relative to the surface normal) along with the original charge compensation system (Kratos Analytical) combining low-energy electrons and electromagnetic field around the sample to effectively compensate charging of the sample. The photoelectrons were collected with the 0° take-off angle (relative to the sample normal) from the spot approximately 1 $mm^2$ with the passing energy of 40 eV. The total energy resolution in these conditions was 0.7 eV (determined from the FWHM of the Ag $3d_{5/2}$ measured at a clean Ag foil). The XPS data were analyzed in the ESCApe software (version 1.6.1.1234). The binding energy was calibrated to 284.8 eV of the C-C component in the C 1s region. All XPS spectra were fitted using the Shirley background and peaks of Gaussian*Lorentzian profile (blend 0.3). The atomic concentration of elements and various compounds was calculated using relative sensitivity factors in the ESCApe built-in database (Kratos Analytical). More details about fitting and atomic concentrations can be found in the Supplementary Materials.

The formal valency of manganese, i. e. the average charge on all Mn atoms in the analyzed sample, was estimated from the atomic concentrations of $Mn^{2+}$ and $Mn^{3+}$ derived from fitting the Mn 2p and Mn3s core level XPS spectra according to:

$$Formal\ Mn\ valency = \frac{At.Conc.(Mn^{2+})*2+At.Conc.(Mn^{3+})*3}{At.Conc.(Mn)}. \quad (3)$$

## 3. Results and discussion

### 3.1 Structural and magnetic properties

XRPD analysis confirmed that our ceramics have the same double perovskite structure with the monoclinic space group $P21/n$ (No. 14, $Z = 2$) as published by Belik [12] – see Figure 1. The lattice parameters are $a$ = 5.31951(2) Å, $b$ = 5.49798(2) Å, $c$ = 7.75735(3) Å, and β = 90.0315(7)°. $Mn^{2+}$ and $W^{6+}$ exhibit a full rock salt-type ordering at the B perovskite sites. In the A-site, $Mn^{2+}$ and $Ca^{2+}$ are statistically distributed. XRPD revealed only one impurity, $CaWO_4$, with a concentration of 3.4%, see Fig. S1 in Supplementary Material (SM).

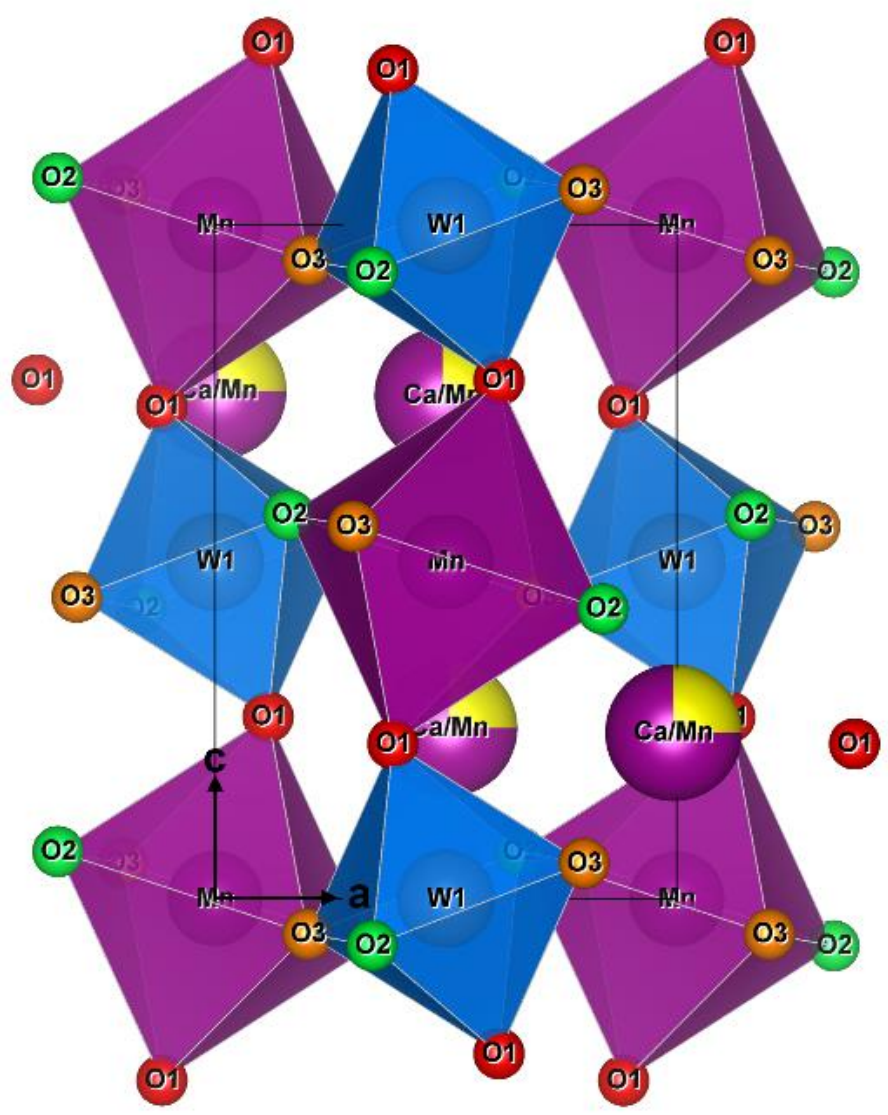

**Figure 1.** Crystal structure of $(Ca_{0.5}Mn_{1.5})MnWO_6$ viewed along the b axis and plotted using Vesta.

We characterized the chemical composition of the ceramics via WDS. The SEM images (Figure S2 in SM) show the microstructure of the ceramics revealing the main $(Ca_{0.5}Mn_{1.5})MnWO_6$ phase but also grains of $CaWO_4$ and $Mn_3O_4$. The last phase was not detected in XRPD due to a strong absorption of X-rays by Mn ions which means, following our XRPD simulations, that its volume concentration is less than 5%.

XPS of Mn 2p and Mn 3s core level spectra revealed that manganese was present in $Mn^{2+}$ and $Mn^{3+}$ oxidation states [16-18]. Mn formal valency in the $(Ca_{0.5}Mn_{1.5})MnWO_6$ sample estimated according to the equation (3) was approximately +2.2 (±0.1). The origin of $Mn^{3+}$ is most probably $Mn_3O_4$ secondary phase which was identified in the XPS spectra along with $CaWO_4$ [16, 19] and also in WDS (see above). Furthermore, the sample contains $Mn^{2+}$-O-$W^{6+}$ bonding states and $Ca^{2+}$ in unusual Ca-O coordination, which are expected in the $(Ca_{0.5}Mn_{1.5})MnWO_6$ material [20-22].

We also analyzed the original $(Ca_{0.5}Mn_{1.5})MnWO_6$ ceramic sample studied in ref. [15]. Its XRD revealed 3.0 wt % of antiferromagnetic MnO impurities (see Fig. S1b in SM and also Fig. S3 for a SEM image). These impurities were confirmed by EDS and WDS analysis, which also detected trace amounts of CaO.

In Fig. 2, the inverse magnetic susceptibility of new $(Ca_{0.5}Mn_{1.5})MnWO_6$ sample is shown as a function of temperature. It exhibits a change of slope below 45 K and a kink at Néel temperature $T_N$ = 18 K. The magnetic susceptibility was fit by the Curie-Weiss law $\chi(T) = C/(T-\theta)$ in the paramagnetic phase and we obtained $C$ = 11.03 (1) emu·K·mol$^{-1}$·Oe$^{-1}$ and $\Theta$ = -222 (2) K

The negative value of the Curie-Weiss temperature confirms that AFM interactions are predominant. Ref. 15 reported a similar magnetic Curie-Weiss temperature: $\Theta$ = -209 (2) K, but somewhat higher Néel temperature $T_N$ = 22 K. It can be explained by different chemical composition of the two samples as discussed above. We also point out that the detected $Mn_3O_4$ impurity phase in the new sample undergoes ferrimagnetic phase transition at 43 K [23], which can be at the origin of the slope change of $\chi^{-1}(T)$ near 45 K. This anomaly is not visible in the previously published sample in ref. [12], because it does not contain $Mn_3O_4$ but small amount of MnO impurity, which has a Neel temperature of 122 K.

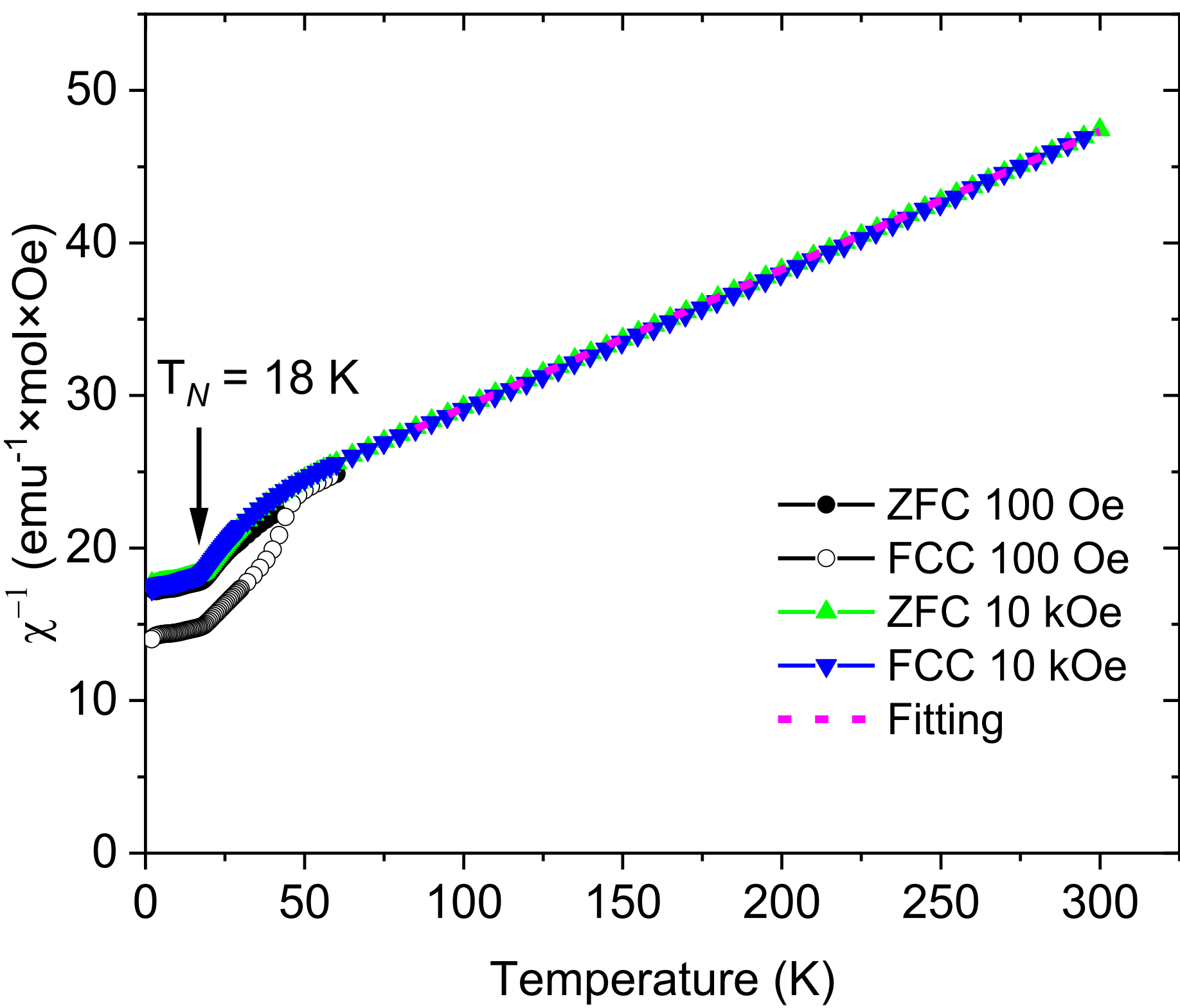

**Figure 2.** Temperature dependence of inverse magnetic susceptibility of $(Ca_{0.5}Mn_{1.5})MnWO_6$ (new sample) measured at two magnetic fields on heating after zero field cooling (ZFC) and field-cooled measured on cooling (FCC). The pink dashed line is the result of the Curie-Weiss fit.

### 3.2 Absence of ferroelectric or antiferroelectric order

To investigate the low-temperature dielectric and polar properties of $(Ca_{0.5}Mn_{1.5})MnWO_6$, broadband dielectric spectroscopy and pyroelectric current measurements were performed. Figure 3 shows the temperature dependence of the real part of the dielectric permittivity, ε', and the dielectric loss tangent, tan δ, measured from 1 Hz to 950 kHz. A strong dielectric dispersion is evident, particularly above 150 K. This is caused by an inhomogeneous conductivity of the grains and grain boundaries and the associated Maxwell-Wagner relaxation similarly as in other slightly conducting ceramics [24-27]. The inhomogeneous conductivity in our ceramics is probably caused by oxygen vacancies. With decreasing temperature, the conductivity decreases, which is why the frequency dependence of $\varepsilon'$ and tanδ also decreases. A small anomaly is observed in the temperature dependence of the permittivity at 18 K, which corresponds to $T_N$. This anomaly is, however, much weaker than typical anomalies accompanying ferroelectric or antiferroelectric phase transitions. Note also that it occurs 4 K lower than in the sample from Ref. [12]. Belik provided us with the original ceramic studied in Ref. [12], and we remeasured its ε'($T$). The obtained dependence is similar to the published one in Ref. [12] (see Fig. S4 in SM) and we also confirm that the original ceramic exhibits an AFM transition at $T_N$ = 22 K. ε'($T$) in [12] exhibits Maxwell-Wagner relaxation only above 200 K. This is probably due to the low conductivity of impurities (MnO and CaO) and the likely lower concentration of oxygen vacancies than in our investigated ceramics.

Since we know that the ceramics we are studying contain 3-4% $Mn_3O_4$ and $CaWO_4$ impurities, it would be appropriate to model our dielectric response as $(Ca_{0.5}Mn_{1.5})MnWO_6$ composite with 3-4% $Mn_3O_4$ and $CaWO_4$. $CaWO_4$ exhibits low permittivity of ~ 8.8 ($\varepsilon'_c$) and ~ 9.5 ($\varepsilon'_a$), which is practically temperature-independent – see Ref. [28]. $Mn_3O_4$ has $\varepsilon'_a \approx 12.4$ and $\varepsilon'_c \approx 17$, which drops down by only approx. 0.05 at $T_N$ [29]. In $Mn_3O_4$ ceramics, the permittivity below 100 K was measured to be even lower ~ 10.2 (probably due to sample porosity), and its change at $T_N$ was even smaller (~0.01) [30]. Unfortunately, since we do not know the exact temperature dependence of the permittivity of the individual phases in our investigated ceramic, we cannot accurately calculate the temperature dependence of the permittivity of the pure $(Ca_{0.5}Mn_{1.5})MnWO_6$ phase. However, the measured permittivity of the studied ceramics is significantly higher (>43) than the permittivity of $Mn_3O_4$ and $CaWO_4$ impurities, and the concentration of impurities is really small (<4%), so it is possible to say that in Fig. 3 we see the dominant dielectric behavior of the $(Ca_{0.5}Mn_{1.5})MnWO_6$ phase.

In $(Ca_{0.5}Mn_{1.5})MnWO_6$ ceramic published in Ref. 12 a small amount (percentage points) of MnO and CaO impurities was detected. MnO has permittivity ~19 at 300 K and it decreases 17% on cooling [31]. Also CaO has relatively low ε' ~ 11.5 [32], so both impurities cannot explain increase of ε' on cooling reported in ref. [12].

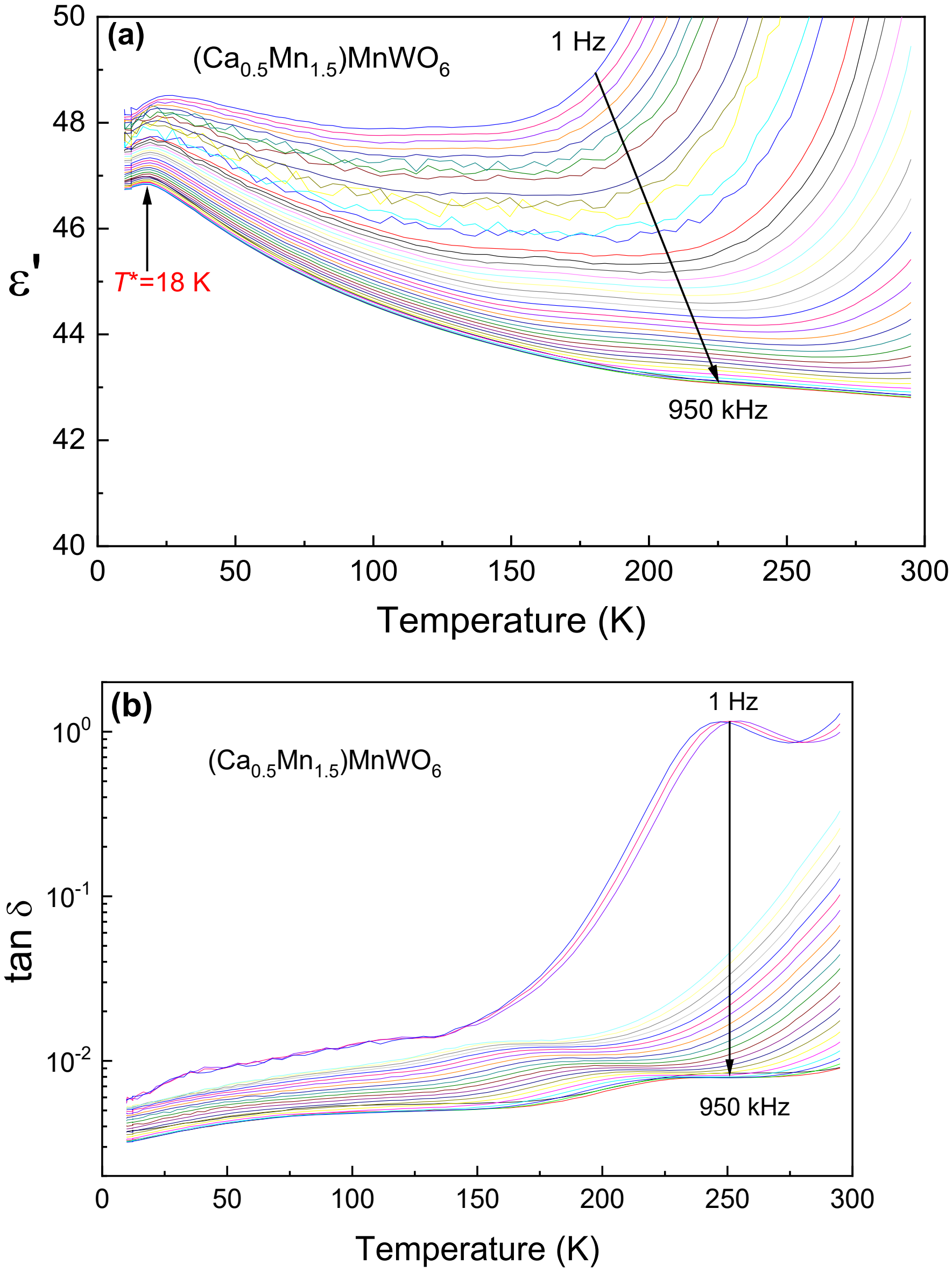

**Figure 3.** Temperature dependence of (a) permittivity $\varepsilon'$ and (b) dielectric loss tanδ measured on cooling over a wide frequency range from 1 Hz to 950 kHz.

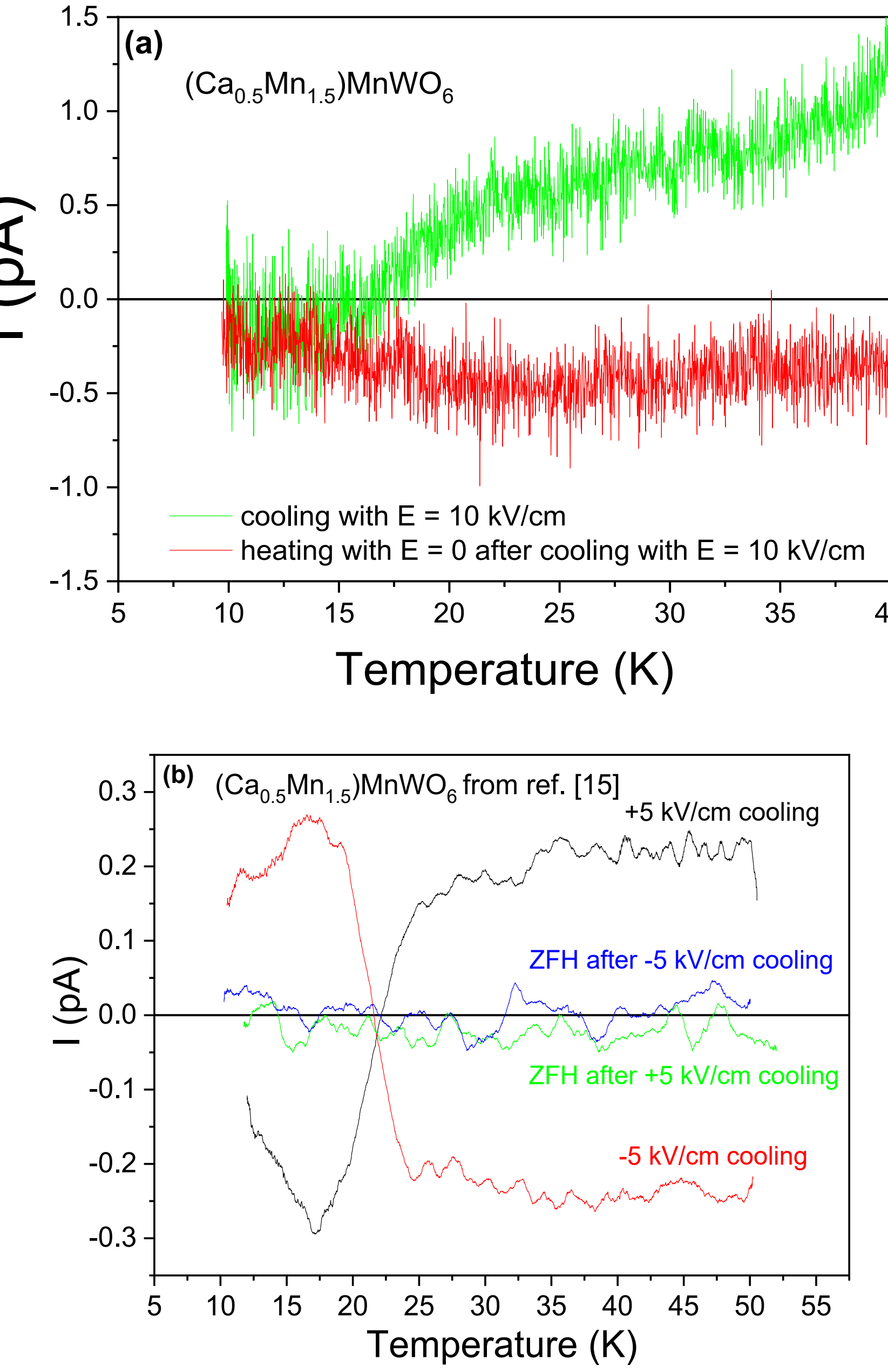

**Figure 4.** The measurement of pyroelectric current of (a) the new $(Ca_{0.5}Mn_{1.5})MnWO_6$ sample during zero-field heating (ZFH) after cooling under an applied field of 10 kV/cm, and (b) the sample from ref. [12] with the poling field of 5 kV/cm. The current measured during the cooling in electric field is also shown.

To examine whether $(Ca_{0.5}Mn_{1.5})MnWO_6$ exhibits a spontaneous electric polarization, pyroelectric current measurements were carried out under poling at 10 kV/cm and short-circuiting protocols. As shown in Fig. 4a, the pyroelectric current measured during zero-field heating does not display any distinguishable peaks or anomalies around $T_N$. The lack of a peak in the pyroelectric current is a strong indication that no long-range ferroelectric ordering is present within this temperature window. The observed signal is dominated by thermally stimulated depolarization currents linked to trapped charges at defects [33]. We also investigated the sample from ref. [12], Fig. 4b. The pyroelectric current measured during a zero-field heating after the cooling in a 5 kV/cm field and a 30-minute dwelling time at 5 K (for relaxation of charged defects) stayed within the noise level and showed no anomaly typical of a ferroelectric phase transition. On the contrary, during the cooling in an electric field, a significant current was observed, which changed the polarity at 22 K. This indicates a migration of charged defects under the electric field. The qualitatively different thermally stimulated depolarization current in both samples seen in Fig. 4 can be explained by different defects in both ceramics.

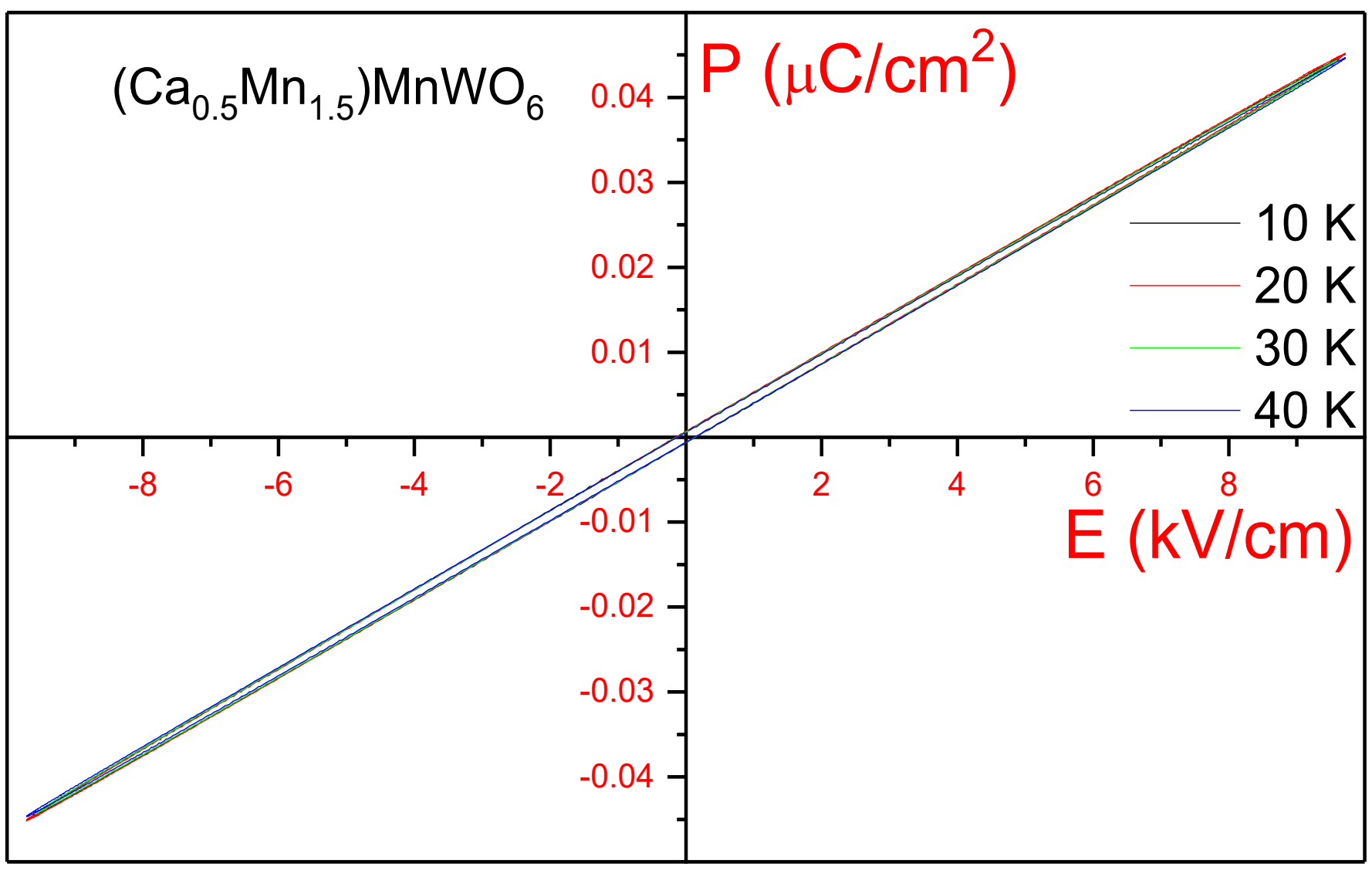

**Figure 5.** Electric field-dependent polarization measured at various temperatures. Only lossy loops are visible.

Additionally, complementary electric field-dependent polarization measurements (P-E hysteresis loop) at various temperatures (10, 20, 30, 40 K) shown in Fig. 5 exhibit almost linear response typical for paraelectrics. Tiny hysteresis is seen at all temperatures, but this is a consequence of the dielectric loss, not of the ferroelectricity. Therefore, this behavior is consistent with a centrosymmetric paraelectric structure of $(Ca_{0.5}Mn_{1.5})MnWO_6$.

Based on his measurements of $\varepsilon'(T)$ and the resulting negative Curie-Weiss temperature, Belik suggested that $(Ca_{0.5}Mn_{1.5})MnWO_6$ could be antiferroelectric below $T_N$ [12], but in the light of our experiments this possibility now seems unlikely. First, the permittivity decrease observed at $T_N$ in Figure 3 is significantly weaker than in ref. [12], and second, we do not see an electric-field-induced transition to the ferroelectric phase in Fig. 5. If the dielectric anomaly observed in $T_N$ were associated with a structural phase transition, it would have to change the IR and Raman selection rules for phonon activity. Therefore, we performed measurements of IR reflectivity, THz transmission, and Raman scattering at different temperatures.

### 3.3 Lattice dynamics and absence of structural phase transition

The room temperature IR reflectivity spectrum of $(Ca_{0.5}Mn_{1.5})MnWO_6$ was measured up to 7000 cm$^{-1}$ and we found that it is completely featureless above 1000 cm$^{-1}$. Therefore the low-temperature IR spectra were measured across the 20-650 cm$^{-1}$ range only, see Fig. 6.

The spectra show a number of reflection bands caused by resonant absorption on polar phonons. The modes above 200 cm$^{-1}$ express bending and stretching vibrations of $MnO_6$ and $WO_6$ octahedra, while lower-frequency modes are more likely related to vibrations of Mn and Ca cations relative to the octahedra.

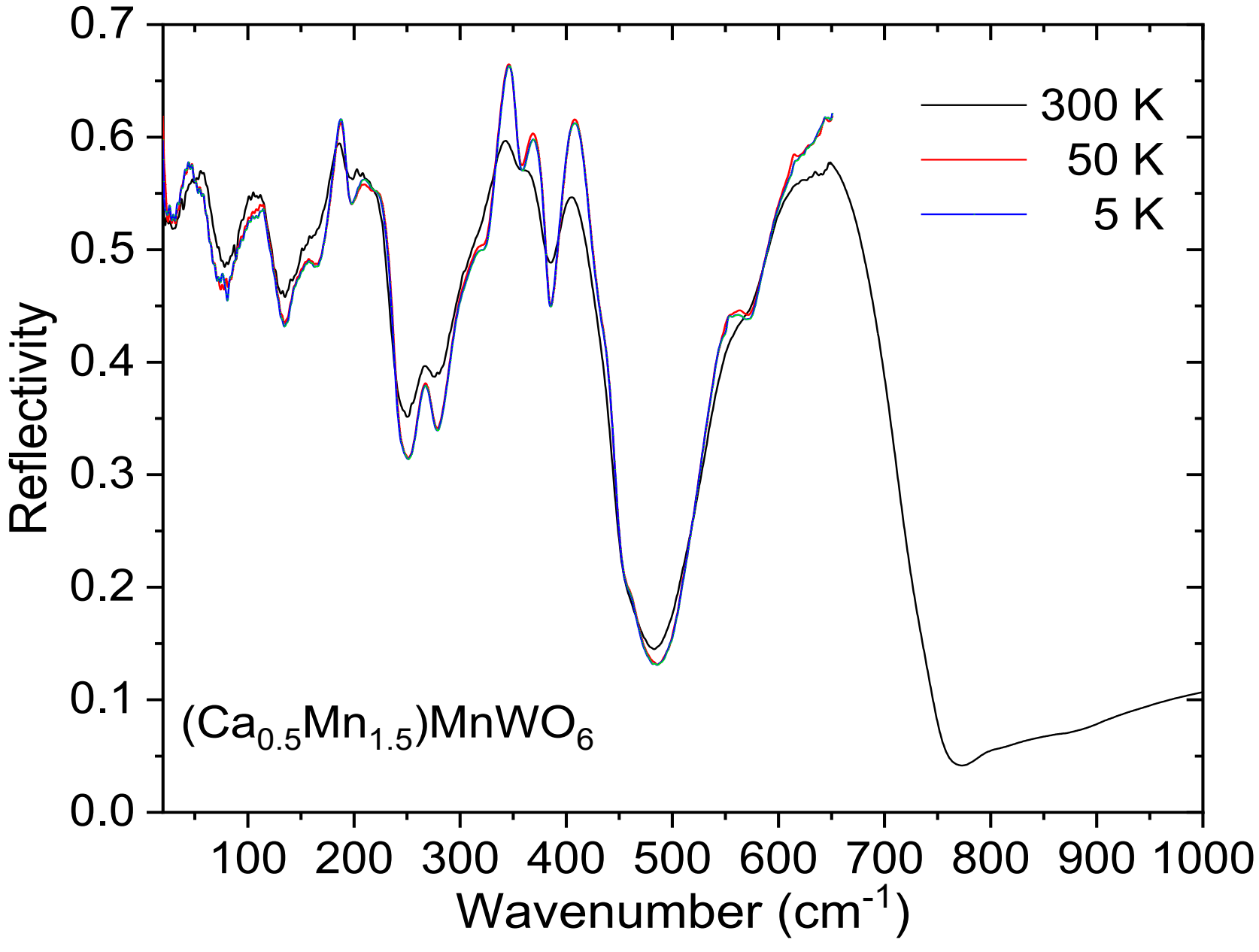

**Figure 6.** IR reflectivity spectra of $(Ca_{0.5}Mn_{1.5})MnWO_6$ at various temperatures.

Upon cooling the sample from 300 K to 5 K, a slight but discernible shift and a sharpening of several phonon modes is observed, particularly near ~270 cm$^{-1}$ and ~350 cm$^{-1}$. These changes indicate a temperature-driven lattice stiffening and a reduced phonon damping at low temperatures. It is also evident that the spectrum does not change significantly between 50 and 5 K (see also Fig. 7 for the spectra in the THz range). This means that the crystal structure does not change near $T_N$ = 18 K.

In principle, the selection rules for IR absorption are quite different in a paraelectric and (anti)ferroelectric phases. In the paraelectric monoclinic $P2_1/n$ phase of $(Ca_{0.5}Mn_{1.5})MnWO_6$, the Mn, W, Ca/Mn, $O_1$, $O_2$, and $O_3$ ions occupy 2$c$, 2$d$, 4$e$ and twice 4$e$ Wyckoff positions, respectively [12]. The factor-group analysis of the $\Gamma$-point optical phonon modes can be expressed by their irreducible representations using tables in [34]:

$$\Gamma_{opt} = 12A_g(x^2, xy) \oplus 18A_u(z) \oplus 12B_g(xz, yz) \oplus 18B_u(x, y) \quad (4)$$

where the $A_g$ and $B_g$ modes are Raman-active, and the $A_u$ and $B_u$ modes are IR-active. Thus in our ceramic we expect 24 Raman-active modes, 33 IR-active modes, and 3 acoustic modes (1$A_u$ and 2$B_u$).

A fit of the IR reflectivity using Eqs. (1) and (2) revealed 23 polar phonons at 5 K (see Table SII in SM for their parameters), which is significantly less than predicted by the factor group analysis. This is typically caused by overlaps of some reflection bands and by a too low intensity of some modes. It would be possible to distinguish the $A_u$ and $B_u$ symmetry modes only in polarized IR spectra of oriented single crystals. Unfortunately, we have ceramics, where the crystal grains are randomly oriented, so it is not possible to distinguish and assign the symmetries of individual modes from our spectra. Note that the IR reflection spectra at 5 and 50 K are identical within the measurement accuracy (see Fig. 6). Thus, we see no changes in phonon frequencies near $T_N$. This is consistent with the saturation of high-frequency $\varepsilon'(T)$ below $T_N$ measured at 950 kHz due to spin-phonon coupling. Permittivity at 1 Hz, on the other hand, shows a slight decrease below $T_N$. This is caused by a change in Maxwell-Wagner relaxation due to decrease of the conductivity below $T_N$.

The sum of the phonon and electron contribution to the permittivity given by the sum of $\varepsilon_\infty + \sum_j \Delta\varepsilon_j$ from Eq. (1) can be directly inferred from the THz spectra of $\varepsilon'(\omega)$ in Fig. 7 (THz dielectric loss spectra $\varepsilon''(\omega)$ are displayed in Fig. S4 in SM) showing the low-frequency wing of the polar phonon spectra. The THz permittivity increases upon cooling, and its values are quite similar to those obtained at 0.95 MHz, see Fig. 3a. This is caused by a small softening of the lowest frequency phonon from 55.6 to 48.4 cm$^{-1}$ with cooling (see Table I is SM). Nevertheless, $\varepsilon'$ is lower than the one obtained in the previous work [12] owing to different amount of defects in the ceramics. It should be noted that we do not see any signature of antiferromagnetic resonance in the THz spectra below $T_N$, meaning that the resonance frequency is probably in the microwave region. This would indicate a weak magnetic anisotropy and a weak internal effective magnetic field in this material. Also, we do not see any new mode to appear in the spectra below $T_{\mathrm{N}}$ expected in (anti)ferroelectric phase with lower symmetry. This confirms a stable crystal structure down to 5 K.

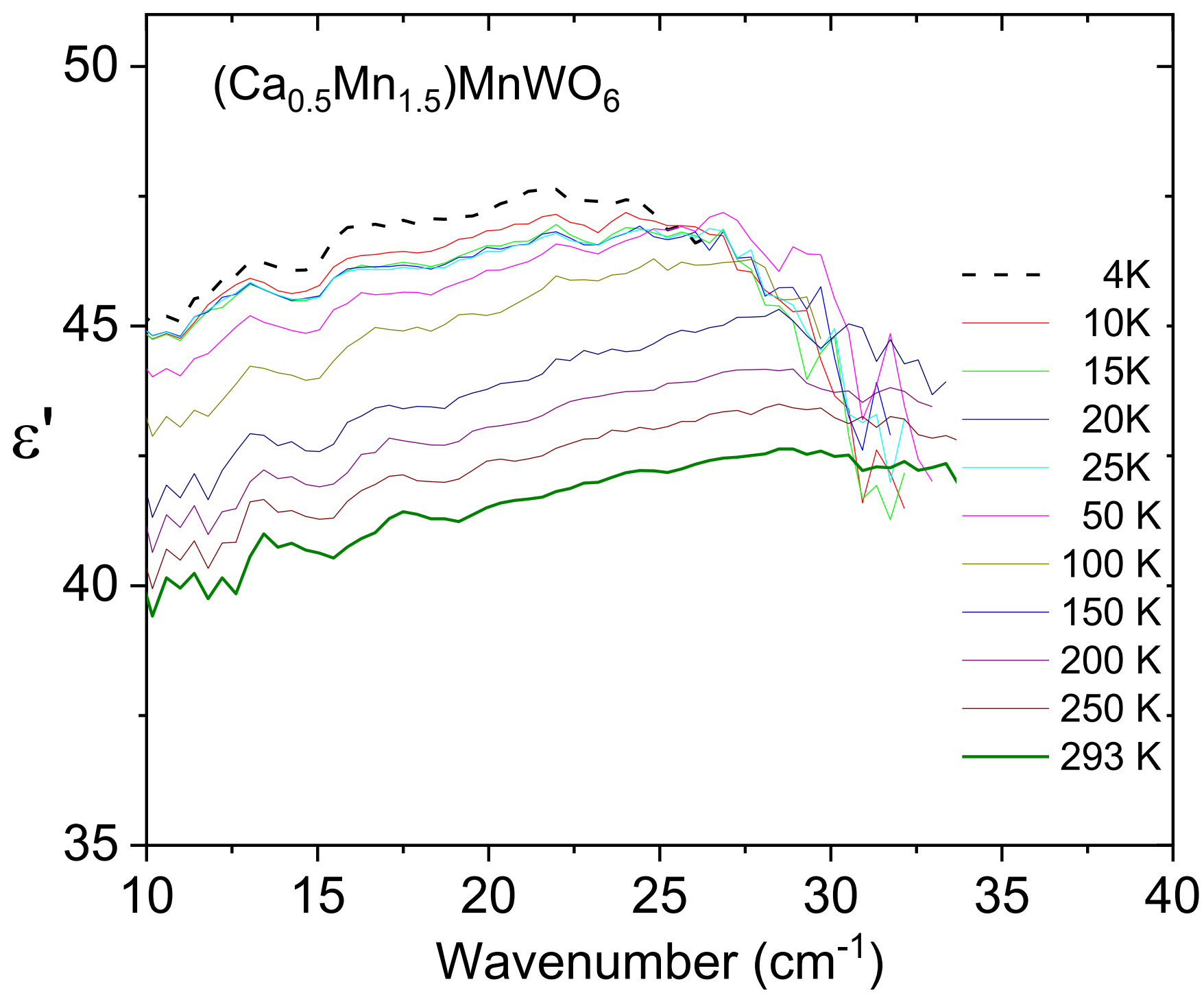

**Figure 7.** Real part of the permittivity of $(Ca_{0.5}Mn_{1.5})MnWO_6$ vs. temperature measured by the THz spectroscopy.

Figure 8 presents the temperature dependent Raman spectra of $(Ca_{0.5}Mn_{1.5})MnWO_6$. The spectra can be divided into two regions. The low-frequency part below 800 cm$^{-1}$ corresponds to single-phonon scattering, while the bands above 1300 cm$^{-1}$ are caused by multi-phonon processes.

Qualitatively similar multi-phonon scattering spectra were observed in $La_{1-x}Sr_xMnO_3$ and $LaMnO_3$ [35], [36].

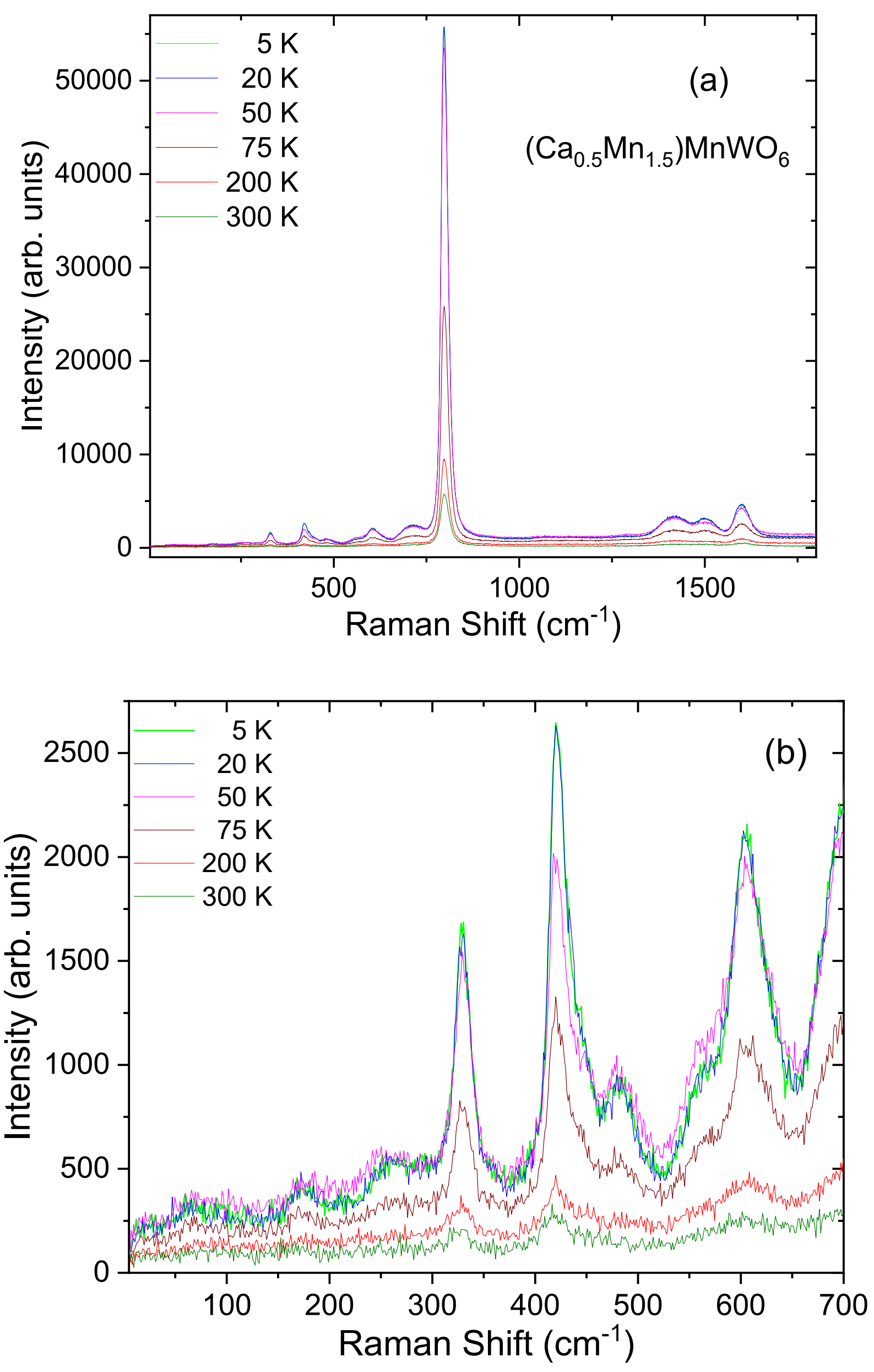

**Figure 8.** (a) Broad-band Raman spectra of $(Ca_{0.5}Mn_{1.5})MnWO_6$ and (b) their low-frequency part.

In the low-frequency part of the spectra, Fig. 8(b), we observe a strong increase of the Raman intensity upon cooling which is quite unusual behavior. The peak height of single-phonon bands usually increases with cooling due to decreasing damping, but in this case, the increase in intensity is much greater. The intensity of multi-phonon scattering above 1300 $cm^{-1}$ also increases, which, on the contrary, should decrease with cooling in two-phonon differential processes due to the smaller population of phonons at low temperatures. Our increase can therefore only be explained by the fact that the material gap changes with temperature and at low temperatures its value approaches the laser energy 2.4 eV and thus we arrive at the resonance enhancement of Raman scattering. Importantly, we do not see any activation of new phonons below $T_N$, which confirms that no structural (antiferroelectric or ferroelectric) phase transition occurs during the magnetic phase transition. Also, no new magnon is activated in the Raman spectra similarly as in THz and IR spectra.

## 4. Conclusion

This study provides a comprehensive characterization of the temperature-dependent physical properties of the double perovskite $(Ca_{0.5}Mn_{1.5})MnWO_6$ ceramics. According to the previous study of $(Ca_{0.5}Mn_{1.5})MnWO_6$ ceramics [12], the material could feature a simultaneous antiferromagnetic and (anti)ferroelectric phase transition at 22 K. However, our measurement of new $(Ca_{0.5}Mn_{1.5})MnWO_6$ ceramic samples prepared in the same laboratory and by the same procedure as in Ref. [12] revealed the Néel temperature at 18 K and only a very weak anomaly of the dielectric permittivity ε'(*T*) at the same temperature. Our further investigations involving also the original ceramics from Ref. [12] show that this anomaly is most likely caused by the spin-phonon coupling and not by a structural change. Pyroelectric studies and *P*(*E*) hysteresis loops measurements did not reveal any signs of spontaneous ferroelectric or antiferroelectric polarization in both new and original samples; the system remains paraelectric at least down to 5 K. THz, IR, and Raman spectra also did not reveal any change in crystal symmetry below $T_N$. Our data therefore do not confirm the conclusions of [12] about the coexistence of AFM and (anti)ferroelectric phases. Nevertheless, we recognize that direct low-temperature structural measurements using X-ray or neutron diffraction would be useful.

In addition, by means of energy dispersion spectroscopy and XPS, we have shown that both new and original samples contain some degree of contamination with secondary phases. These secondary phases are different for the two samples (3.0 wt % of antiferromagnetic MnO and traces of CaO in the original sample; 3.4 wt. % of $CaWO_4$ and a smaller amount of $Mn_3O_4$ in the new sample) and they are thus likely responsible for the different $T_N$ values and different dielectric properties.

## CRediT authorship contribution statement

**Hong Dang Nguyen**: Writing – original draft, Investigation, Methodology, Formal analysis, Software, Data curation. **Alexei A. Belik**: Methodology, Data curation, Formal analysis. **Petr Kužel**: Investigation, Writing – review & editing. **Fedir Borodavka**: Investigation, Formal analysis. **Maxim Savinov**: Methodology, Investigation, Formal analysis. **Klára Beranová:** Methodology, Investigation, Formal analysis. **Petr Proschek**: Methodology, Investigation, Formal analysis. **Jan Drahokoupil:** Methodology, Investigation, Formal analysis. **Markéta Jarošová**: Methodology, Investigation, Formal analysis. **Bartoloměj Vaníček:** Investigation **Stanislav Kamba**: Writing – review & editing, Validation, Project administration, Investigation, Funding acquisition, Supervision.

## Declaration of Competing Interest

The authors declare that they have no known competing financial interests or personal relationships that could have appeared to influence the work reported in this paper.

## Acknowledgements

This work was supported by the Czech Science Foundation (Project No. 24-10791S) and by the project TERAFIT – CZ.02.01.01/00/22_008/0004594 co-financed by the European Union and the Ministry of Education, Youth and Sports of the Czech Republic. We would like to thank M. Vondráček for his help with the XPS measurements.

## Data availability

Data will be made available on request.

# Supporting Materials

## Dielectric, magnetic and lattice dynamics properties of double perovskite $(Ca_{0.5}Mn_{1.5})MnWO_6$

*Hong Dang Nguyen*[1,2,*], *Alexei A. Belik*[3], *Petr Kužel*[1], *Fedir Borodavka*[1], *Maxim Savinov*[1], *Klára Beranová*[4], *Jan Drahokoupil*[1], *M. Jarošová*[1], *Petr Proschek*[5], *Bartoloměj Vaníček*[1], *Stanislav Kamba*[1,*]

[1]*Institute of Physics, Czech Academy of Sciences, Na Slovance 2, 182 00 Prague 8, Czech Republic*

[2]*Faculty of Nuclear Sciences and Physical Engineering, Czech Technical University in Prague, Břehová 7, 115 19 Prague 1, Czech Republic*

[3]*Research Center for Materials Nanoarchitectonics (MANA), National Institute for Materials Science (NIMS), Namiki 1-1, Tsukuba, Ibaraki 305-0044, Japan*

[4]*Institute of Physics, Czech Academy of Sciences, Cukrovarnická 10, 162 00 Prague 6, Czech Republic*

[5]*Department of Condensed Matter Physics, Faculty of Mathematics and Physics, Charles University, Ke Karlovu 5, Prague 2, 121 16 Czech Republic*

[*]Corresponding authors: kamba@fzu.cz, dang@fzu.cz

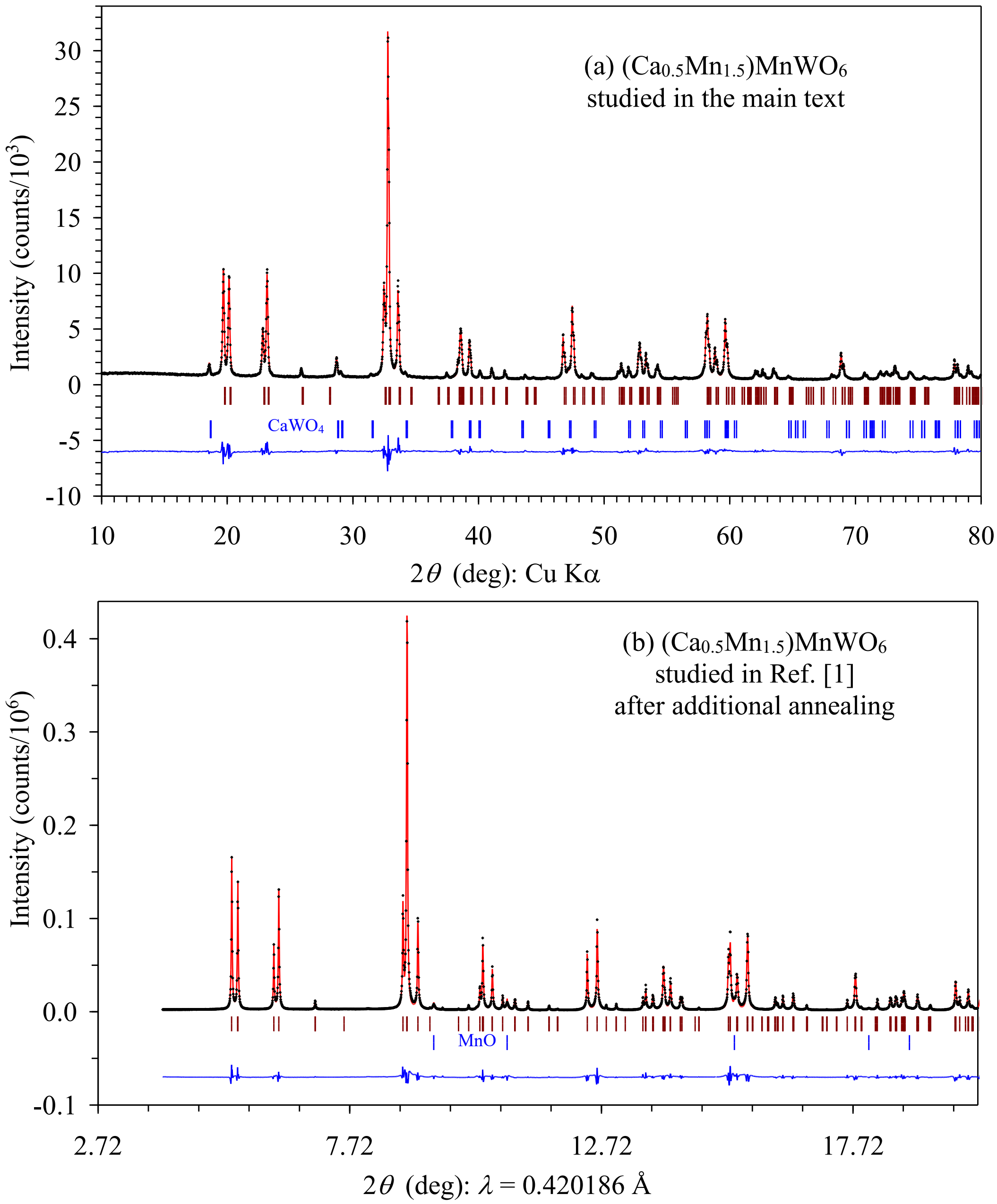

**Figure S1.** Room-temperature experimental (black dots), calculated (red lines), and difference (blue lines on the bottom) X-ray diffraction data of $(Ca_{0.5}Mn_{1.5})MnWO_6$ ceramics investigated in the main text and measured using laboratory X-ray equipment (a) and in Ref. [1] after an additional high-pressure high-temperature annealing and measured using a synchrotron source (b). The tick marks show possible Bragg reflections for the main phases (the first row) and impurities (the second row) ($CaWO_4$ (about 3.4 wt. %) in (a) and MnO in (b) (about 3.6 wt. %)). The sample on panel (b) was annealed at 6 GPa and 1550 K for 2h in an Au capsule as in Ref. [1] plus at 6 GPa and 1900 K for 1h in a Pt capsule.

**Table SI.** Refined structural parameters of $(Ca_{0.5}Mn_{1.5})MnWO_6$ (studied in Ref. [1], after an additional annealing) at room temperature from synchrotron powder X-ray diffraction data [a]

| Site | Wyck. | $x$ | $y$ | $z$ | $B_{iso.}$ (Å$^2$) |
|---|---|---|---|---|---|
| Mn | 2*c* | 0.5 | 0.0 | 0.5 | 0.65(2) |
| W | 2*d* | 0.5 | 0.0 | 0.0 | 0.489(8) |
| Ca/Mn | 4*e* | 0.9936(9) | 0.0491(3) | 0.2461(3) | 0.92(3) |
| O1 | 4*e* | 0.3816(11) | 0.9435(13) | 0.2312(12) | 0.84(17) |
| O2 | 4*e* | 0.1631(15) | 0.2082(15) | 0.5742(14) | 1.66(23) |
| O3 | 4*e* | 0.7087(15) | 0.3293(15) | 0.4458(14) | 1.29(21) |

[a] Space group $P2_1/n$ (No. 14, cell choice 2), $Z = 2$. Wavelength: $\lambda = 0.420186$ Å. Occupation factors of the Mn, W, O1, O2, and O3 sites are unity ($g$ = 1); the occupation factor of the Ca/Mn site is 0.25Ca + 0.75Mn. Wyck.: Wyckoff position.

$a$ = 5.31220(4) Å, $b$ = 5.49141(5) Å, $c$ = 7.75726(6) Å, $\beta$ = 90.0549(11)°, and $V$ = 226.291(3)Å$^3$; $\rho$ = 6.417 g/cm$^3$; $R_{wp}$= 8.58 %, $R_p$= 6.19 %, $R_B$ = 3.73 %, and $R_F$ = 1.90 %. Impurities: MnO (3.6 wt. %).

Note that CaO detected by EDS could become amorphous and could not be detected by XRD. Note that additional refinements of the occupation factors of the Mn and W sites gave the following values, $g$(Mn) = 0.995(4) and $g$(W) = 0.988(3). That is, the values were close to 1 within the accuracy of the method confirming the full rock-salt B-site ordering; therefore, they were fixed to 1 in the final model.

Synchrotron XRPD data were collected at room temperature on the beamline BL02B2 [2] of SPring-8 (the intensity data were taken between 1.95° and 78.09° at 0.006° intervals in 2θ using a wavelength of λ = 0.420186 Å). The data between 4.00° and 70.00° were used in the refinement. The sample was placed into an open Lindemann glass capillary tube (inner diameter: 0.2 mm), which was rotated during measurements. The Rietveld analysis was performed using the *RIETAN-2000* program [3].

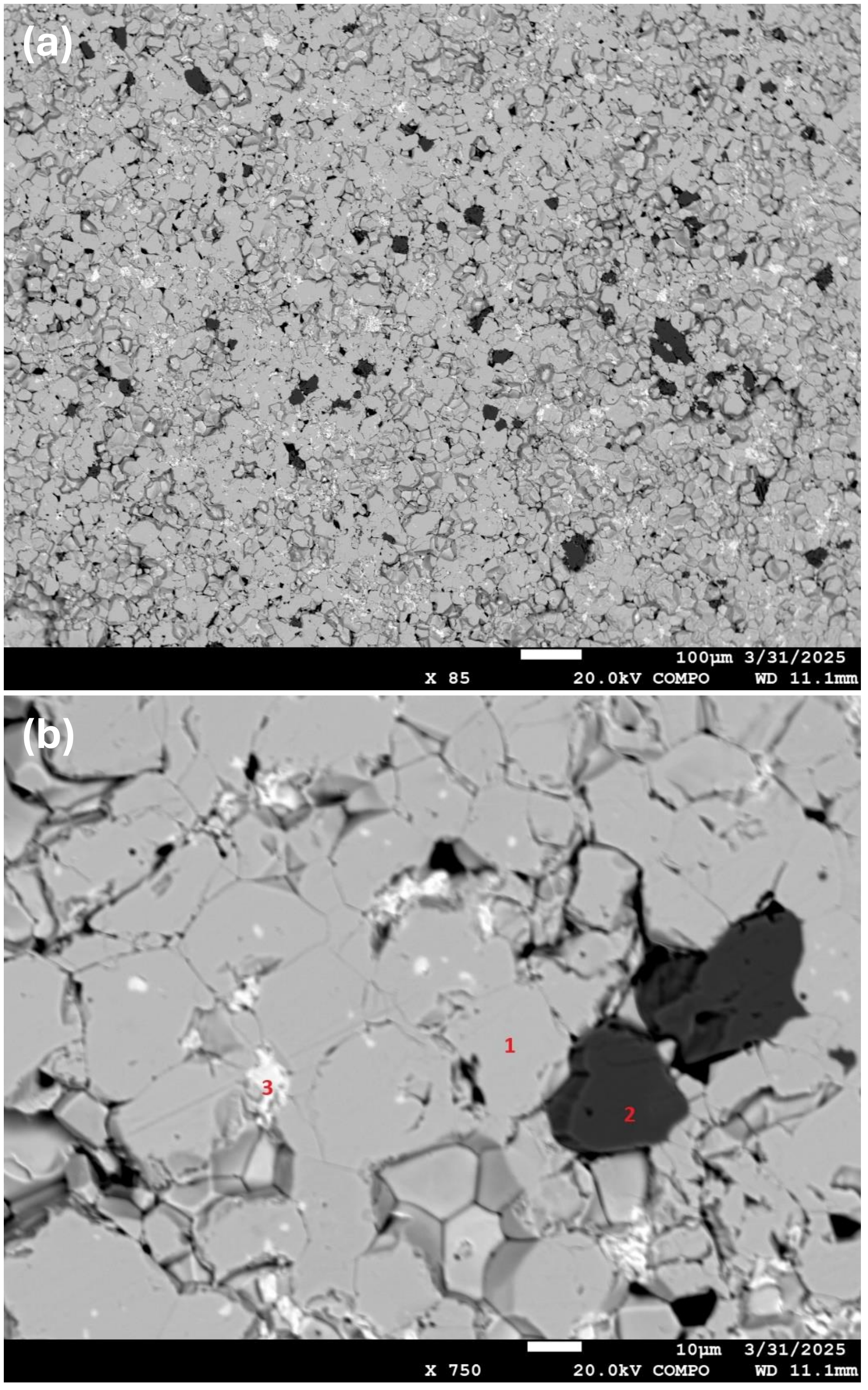

**Figure S2.** (a-b) SEM images (BSE mode) of $(Ca_{0.5}Mn_{1.5})MnWO_6$ ceramics microstructure. Points labeled (1), (2), and (3) are the locations where WDS analysis was carried out. Large grey areas correspond to $(Ca_{0.5}Mn_{1.5})MnWO_6$ , white areas are $CaWO_4$ and black grains are $Mn_3O_4$.

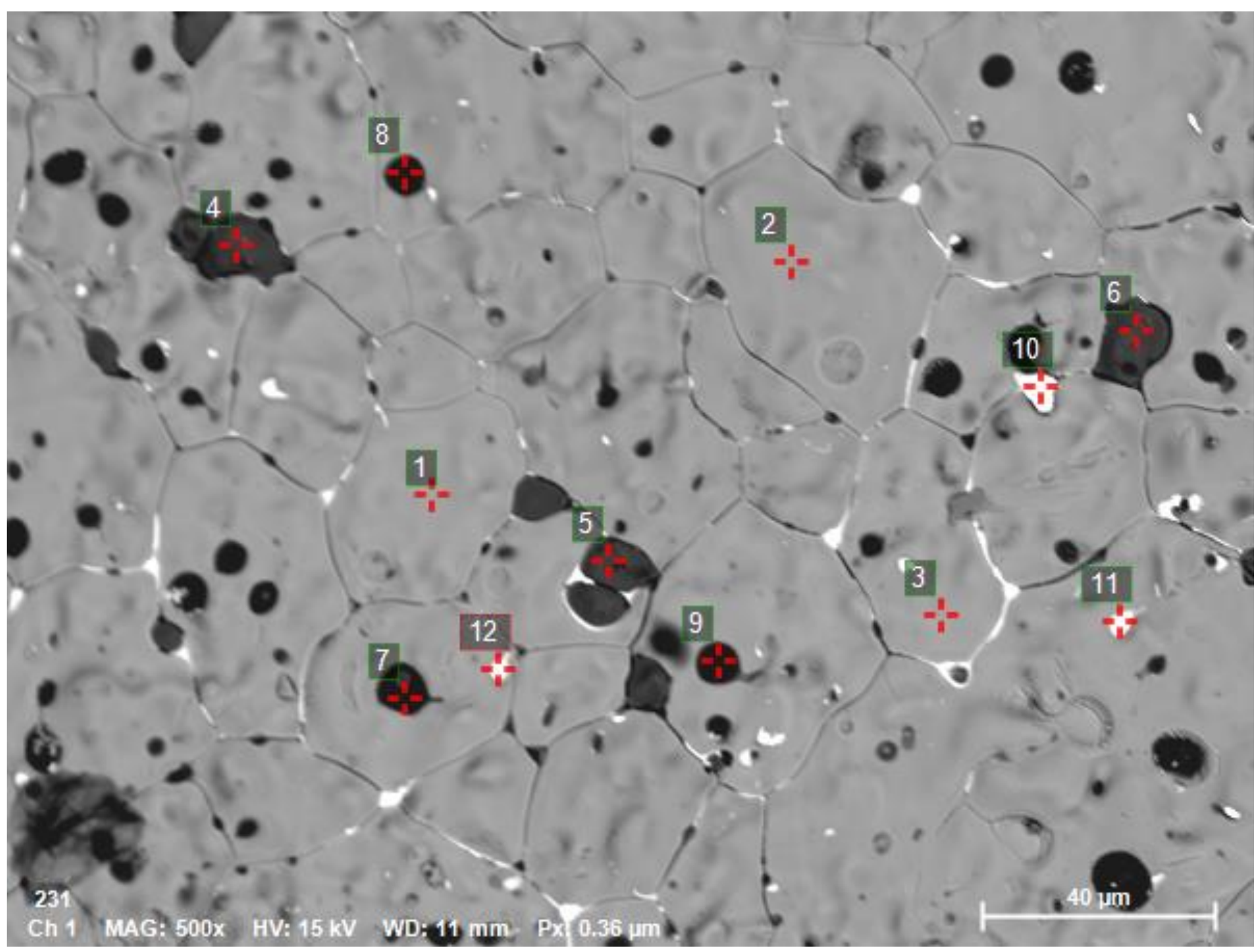

**Figure S3**. SEM images (BSE mode) of the ceramics published in ref. [1]. The numbers indicate the points where EDS and WDS analysis were performed. Large grey areas labeled 1-3 correspond to $(Ca_{0.5}Mn_{1.5})MnWO_6$ , black grains (marked 4-6) show MnO, other black areas marked 7-9 are CaO+MnO and white spots 10-12 are the rest of Pt electrodes. The sample was annealed at 6 GPa and 1550 K for 2h in an Au capsule as in Ref. [1] plus at 6 GPa and 1900 K for 1h in a Pt capsule.

$CaWO_4$ impurity was observed by XRD after the 6 GPa-1550 K annealing, which disappeared after the 6 GPa-1900 K additional annealing. Magnetic and dielectric measurements confirmed that magnetic and dielectric anomalies were nearly identical after the 6 GPa-1550 K annealing and after the 6 GPa-1900 K additional annealing.

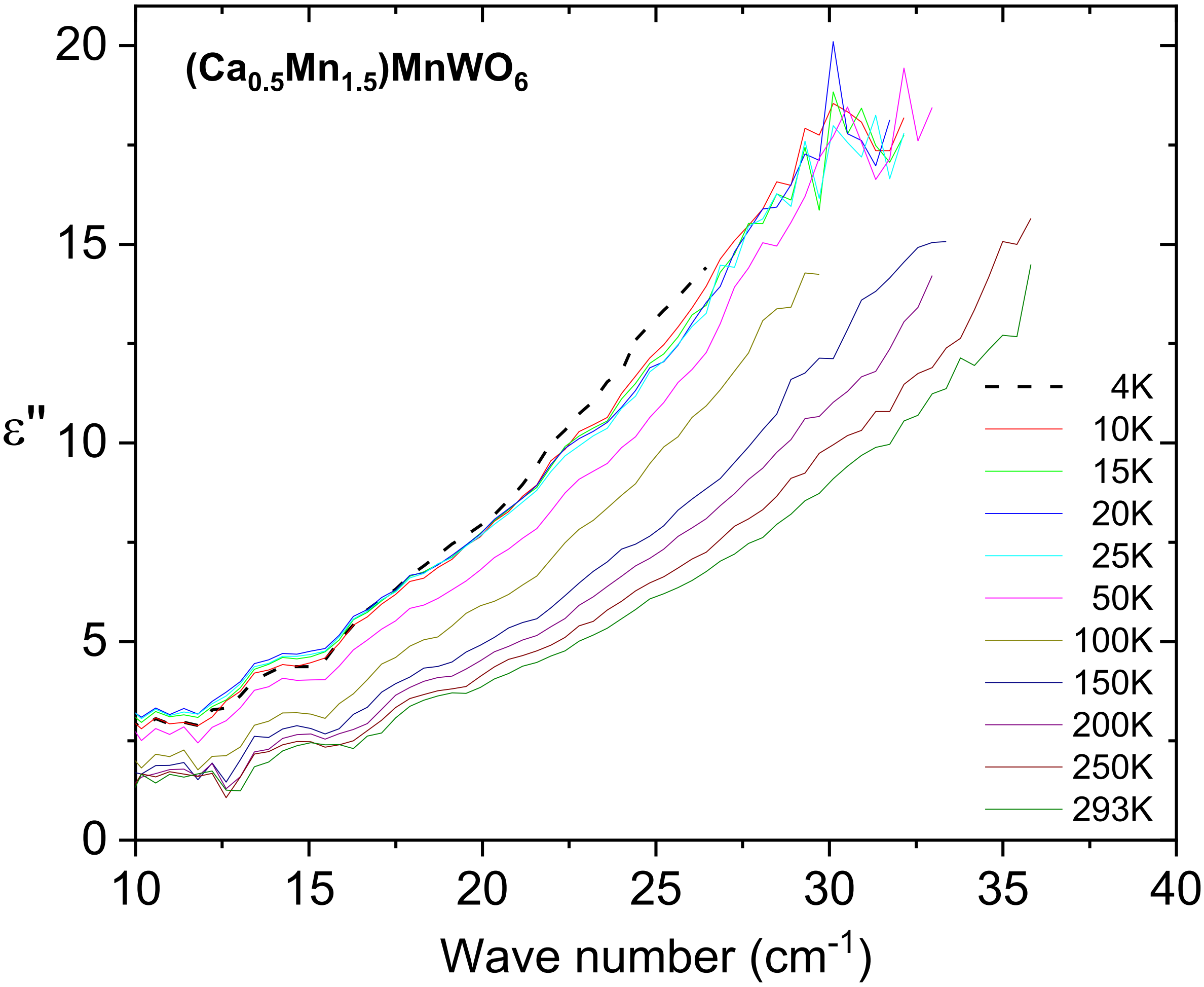

**Figure S4.** Temperature-dependence of imaginary part of permittivity $\varepsilon''$ in $(Ca_{0.5}Mn_{1.5})MnWO_6$. The increase in losses with cooling is caused by the softening of the polar phonon above 50 cm$^{-1}$, visible in Fig. 6.

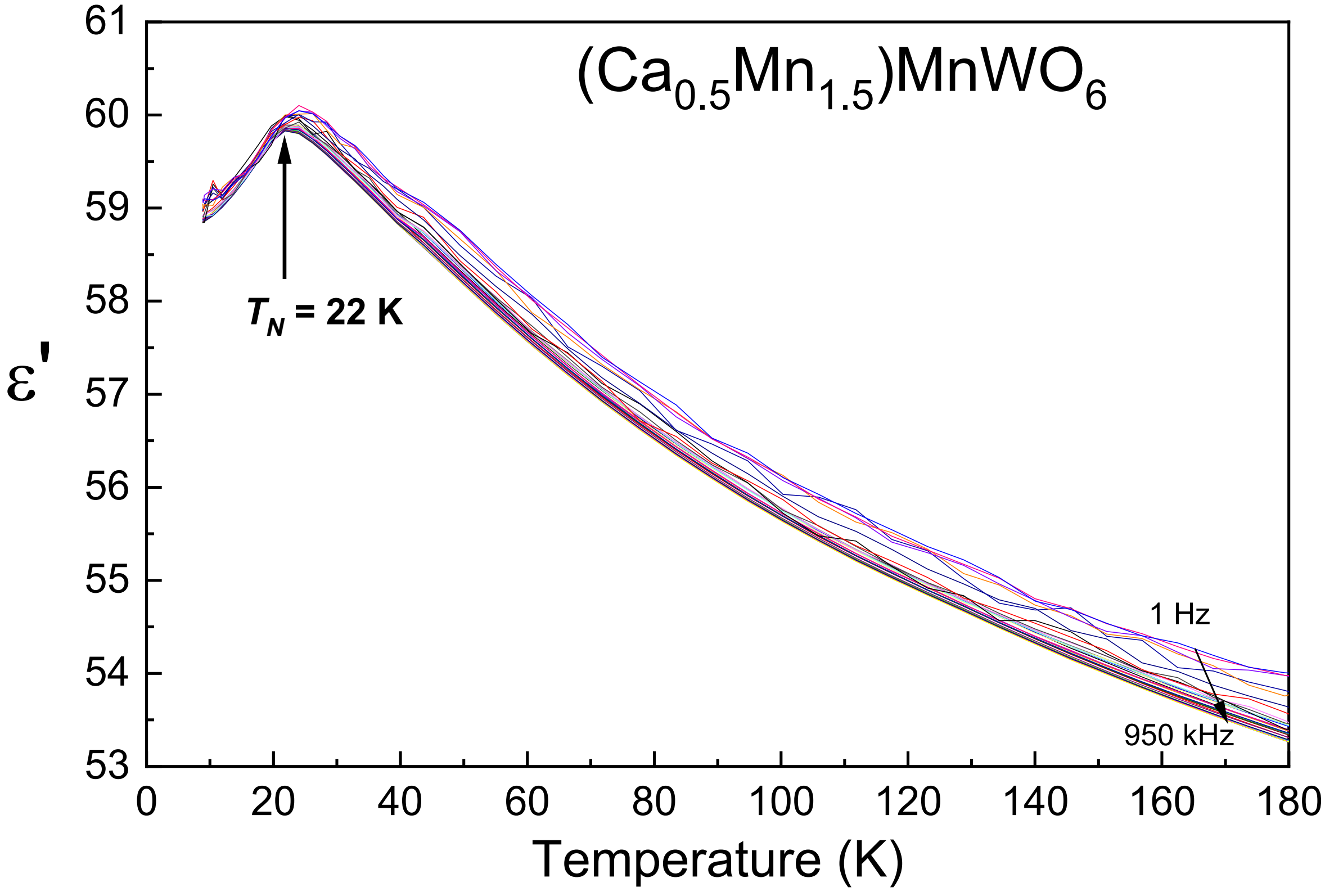

**Figure S5.** Repeated measurement of the temperature dependence of dielectric permittivity (1 Hz – 950 kHz) in a sample of $(Ca_{0.5}Mn_{1.5})MnWO_6$ published in ref. [1]. The sample was annealed at 6 GPa and 1550 K for 2h in an Au capsule as in Ref. [1] plus at 6 GPa and 1900 K for 1h in a Pt capsule.

**TABLE SII.** Room-temperature and 5 K phonon parameters obtained from the optimum reflectance fit by Lorentzian model with $\varepsilon_\infty$ = 2.6. Parameters of the fit are the same at 5 K and 50 K (not shown), so the structure does not change in this temperature range. The smaller number of phonons distinguished at 300 K is caused by higher phonon damping, which prevents some modes from being distinguished.

| Temperature | No. j | $\omega_{TOj}$ (cm$^{-1}$) | $\gamma_j$ (cm$^{-1}$) | $\Delta\varepsilon_j$ |
|---|---|---|---|---|
| 300 K | 1 | 55.6 | 32.6 | 22.4 |
| | 2 | 101.3 | 43.1 | 7.7 |
| | 3 | 155 | 47.3 | 2.3 |
| | 4 | 179.8 | 23.4 | 0.6 |
| | 5 | 208.9 | 49.8 | 0.8 |
| | 6 | 265.5 | 32.2 | 0.2 |
| | 7 | 306.7 | 56.2 | 0.6 |
| | 8 | 336 | 41.2 | 0.3 |
| | 9 | 365.3 | 38.2 | 0.15 |
| | 10 | 399.1 | 38.5 | 0.17 |
| | 11 | 431.8 | 56.3 | 0.11 |
| | 12 | 553.6 | 59.4 | 0.14 |
| | 13 | 611.1 | 135.1 | 0.6 |
| | 14 | 677.6 | 125 | 0.01 |
| 5 K | 1 | 48.4 | 13.4 | 4.2 |
| | 2 | 60 | 20.6 | 5.5 |
| | 3 | 77.1 | 8.4 | 0.5 |
| | 4 | 106.5 | 42.2 | 4.7 |
| | 5 | 117 | 21.9 | 1.8 |
| | 6 | 158.4 | 44.7 | 2.7 |
| | 7 | 185.5 | 16.8 | 2.53 |
| | 8 | 208.3 | 38.1 | 2.71 |
| | 9 | 225 | 29.2 | 0.92 |
| | 10 | 267.2 | 23.1 | 0.43 |
| | 11 | 301 | 7.4 | 0.05 |
| | 12 | 315 | 45.9 | 1.94 |
| | 13 | 340.5 | 23.2 | 2.0 |
| | 14 | 365.3 | 27.3 | 0.6 |
| | 15 | 397.4 | 23 | 0.7 |
| | 16 | 416.6 | 37.6 | 0.3 |
| | 17 | 434.1 | 24.1 | 0.08 |
| | 18 | 447 | 17.4 | 0.1 |
| | 19 | 500 | 2.1 | 0.002 |
| | 20 | 562.6 | 45.5 | 0.95 |
| | 21 | 615.2 | 136.2 | 0.6 |
| | 22 | 680.2 | 115 | 0.1 |
| | 23 | 700 | 37.1 | 4.04 |

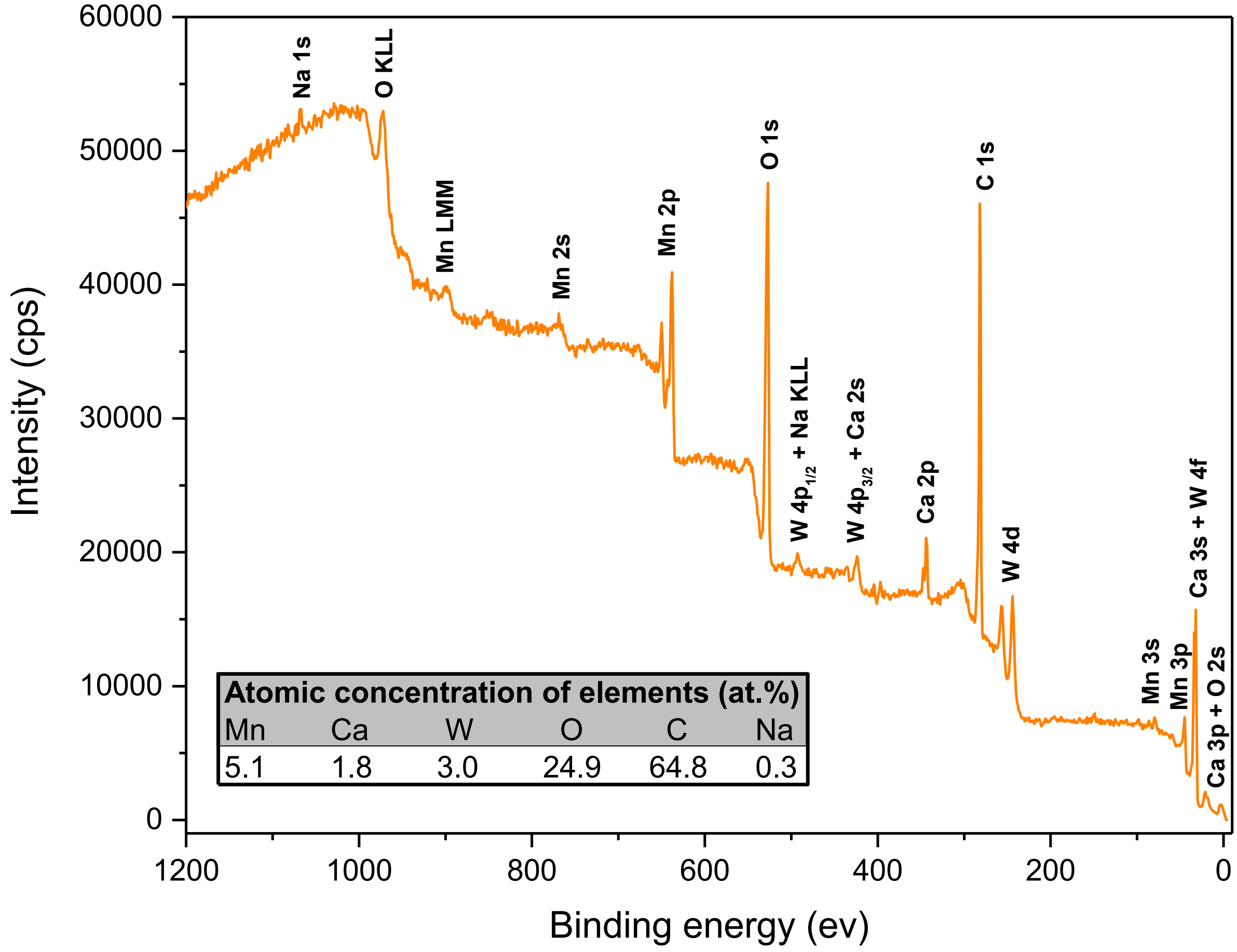

**Figure S6.** The XPS survey scan of the powdered ($Ca_{0.5}Mn_{1.5}$)$MnWO_6$ with indicated identified photoelectron lines. Atomic concentration of elements was calculated from the Mn $2p_{3/2}$, Ca 2p, W 4f, O 1s, C 1s and Na 1s core levels using the ESCApe software (Kratos, version 1.6.1.1234) and its built-in library of the relative sensitivity factors.

**TABLE SIII.** Atomic concentrations of all identified compounds in the powdered ($Ca_{0.5}Mn_{1.5}$)$MnWO_6$ calculated from the fitted XPS core-level regions.

| **Atomic concentration (at.%)** | | | | | | | | | | | | |
|---|---|---|---|---|---|---|---|---|---|---|---|---|
| **Mn** | | | **Ca** | | **W** | | **O** | | | | **C** | **Na** |
| $Mn^{2+}$ | $Mn^{2+}$ | $Mn^{3+}$ | $Ca^{2+}$ | $Ca^{2+}$ | $W^{6+}$ | $W^{6+}$ | Mn-O-W | Ca-O-W | OH | $H_2O$ | | |
| $CaMnWO_6$ | $Mn_3O_4$ | | $CaMnWO_6$ | $CaWO_4$ | $CaMnWO_6$ | $CaWO_4$ | $CaMnWO_6$ $Mn_3O_4$ | $CaWO_4$ | | | | |
| 4.0 | 0.4 | 0.7 | 0.6 | 1.2 | 2.2 | 0.8 | 8.0 | 1.3 | 12.7 | 2.9 | 64.8 | 0.3 |

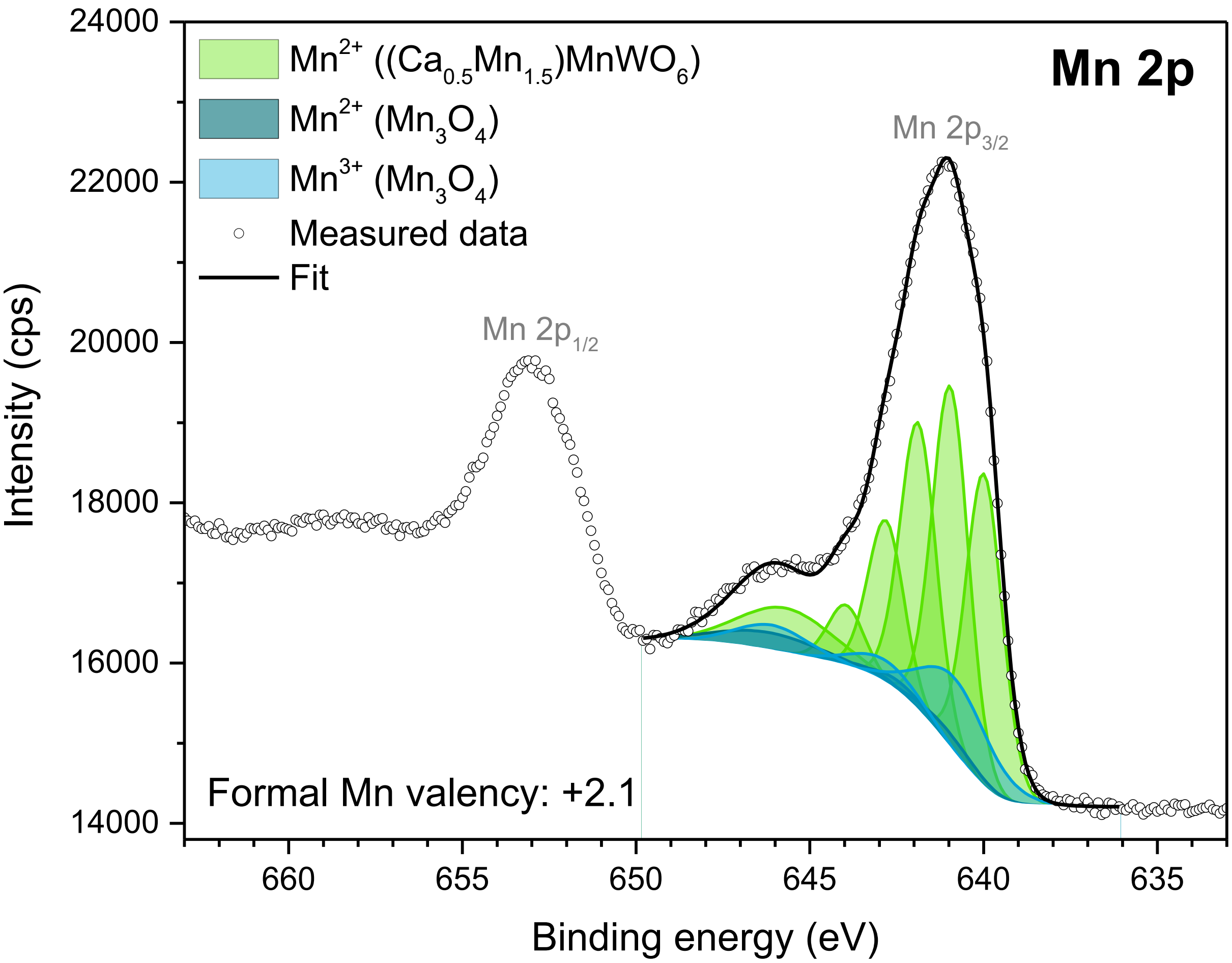

**Figure S7.** The XPS Mn 2p core level region of the powdered $(Ca_{0.5}Mn_{1.5})MnWO_6$. The Mn $2p_{3/2}$ was fitted by Shirley background and three sets of multiplets corresponding to $Mn^{2+}$ from $(Ca_{0.5}Mn_{1.5})MnWO_6$ – green (approximated by $Mn^{2+}$ from MnO [4]), and $Mn^{2+}$ and $Mn^{3+}$ from $Mn_3O_4$ – dark and light blue, respectively [5]. The relative positions, intensities and FWHMs were constrained in each multiplet according to [4, 5]. Formal Mn valency was determined to be +2.1.

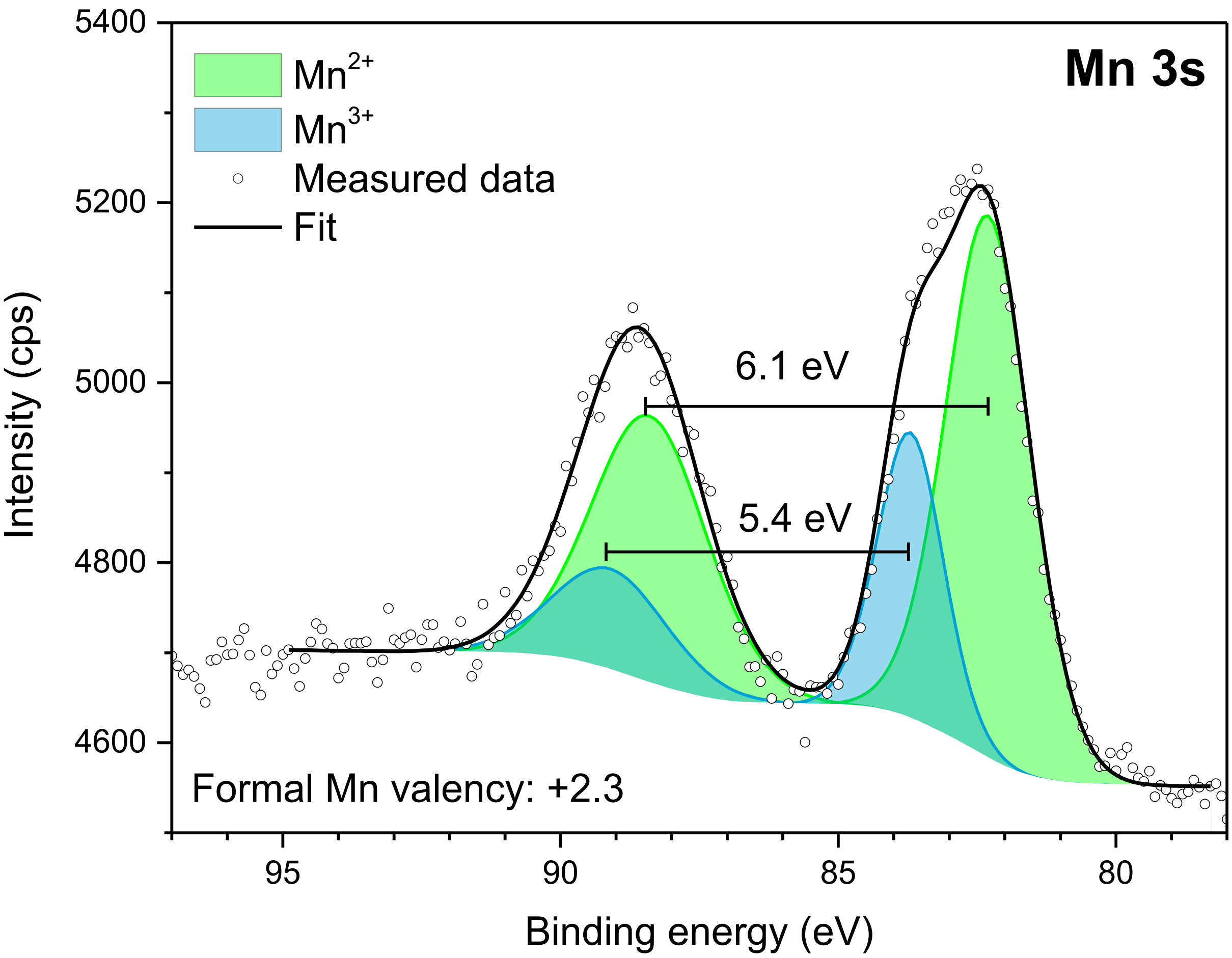

**Figure S8.** The XPS Mn 3s core level region of the powdered $(Ca_{0.5}Mn_{1.5})MnWO_6$. The Mn 3s was fitted by Shirley background and two sets of multiplets corresponding to $Mn^{2+}$ – green and $Mn^{3+}$ – blue [6]. The multiplet splitting and relative intensity ratio in each multiplet was constrained according to [6]. Formal Mn valency was determined to be +2.3.

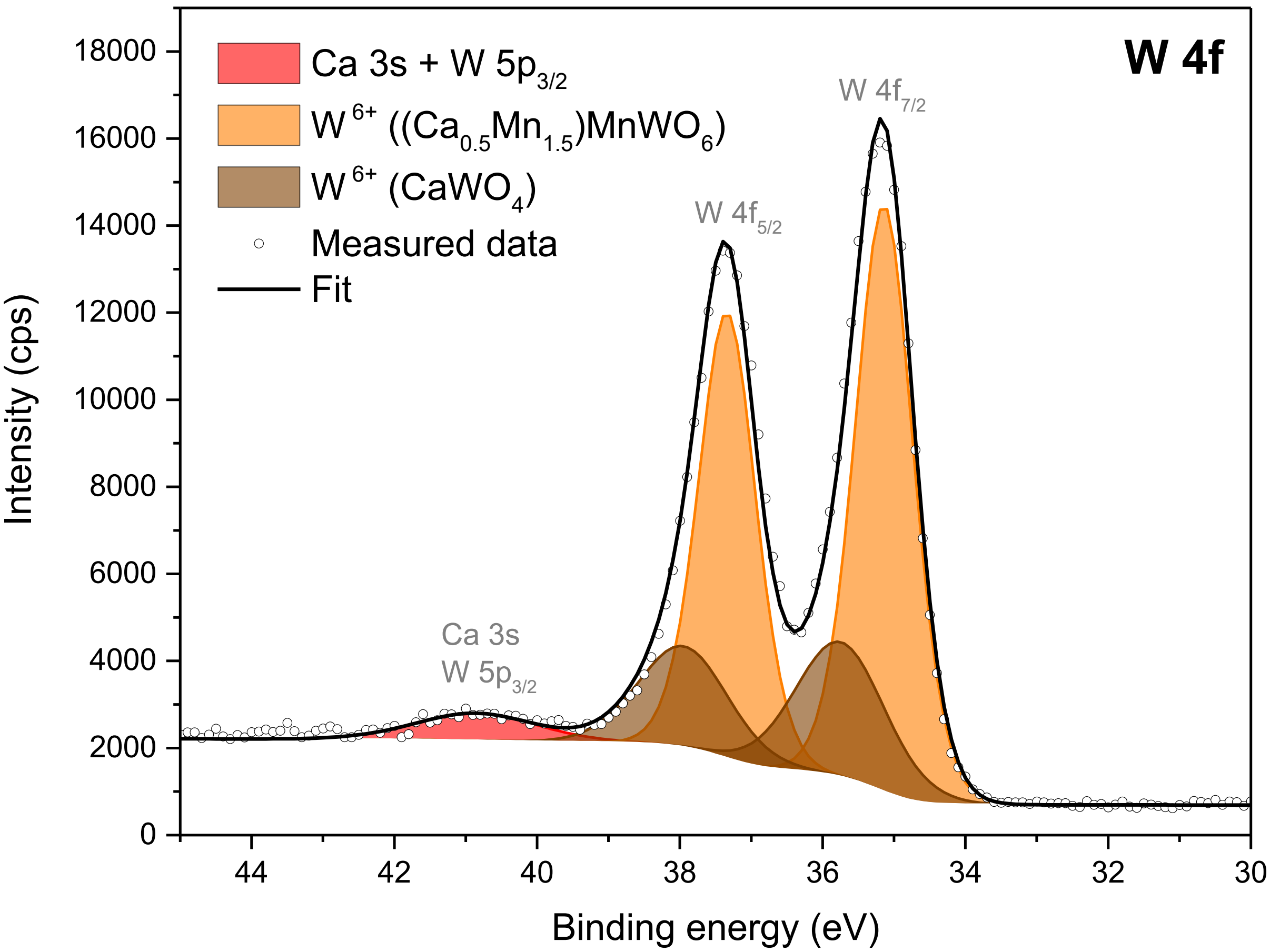

**Figure S9.** The XPS W 4f core level region of the powdered ($(Ca_{0.5}Mn_{1.5})MnWO_6$). The W 4f was fitted by Shirley background and two sets of spin-orbit $4f_{7/2}$ – $4f_{5/2}$ doublets. The binding energy of the orange doublet (35.1 – 37.3 eV) is very similar to W 4f from $MnWO_4$ [7, 8]. Therefore, the orange doublet was attributed to $W^{6+}$ from the $(Ca_{0.5}Mn_{1.5})MnWO_6$. The brown W 4f doublet (35.8 – 38.0 eV) corresponds to $W^{6+}$ from $CaWO_4$ [9]. The spin-orbit splitting and relative intensity ratio in each doublet was fixed to 2.2 eV and 3:4, respectively [10]. An overlapping signal from Ca 3s and W $5p_{3/2}$ is approximately 5.5 eV above W $4f_{7/2}$ [10].

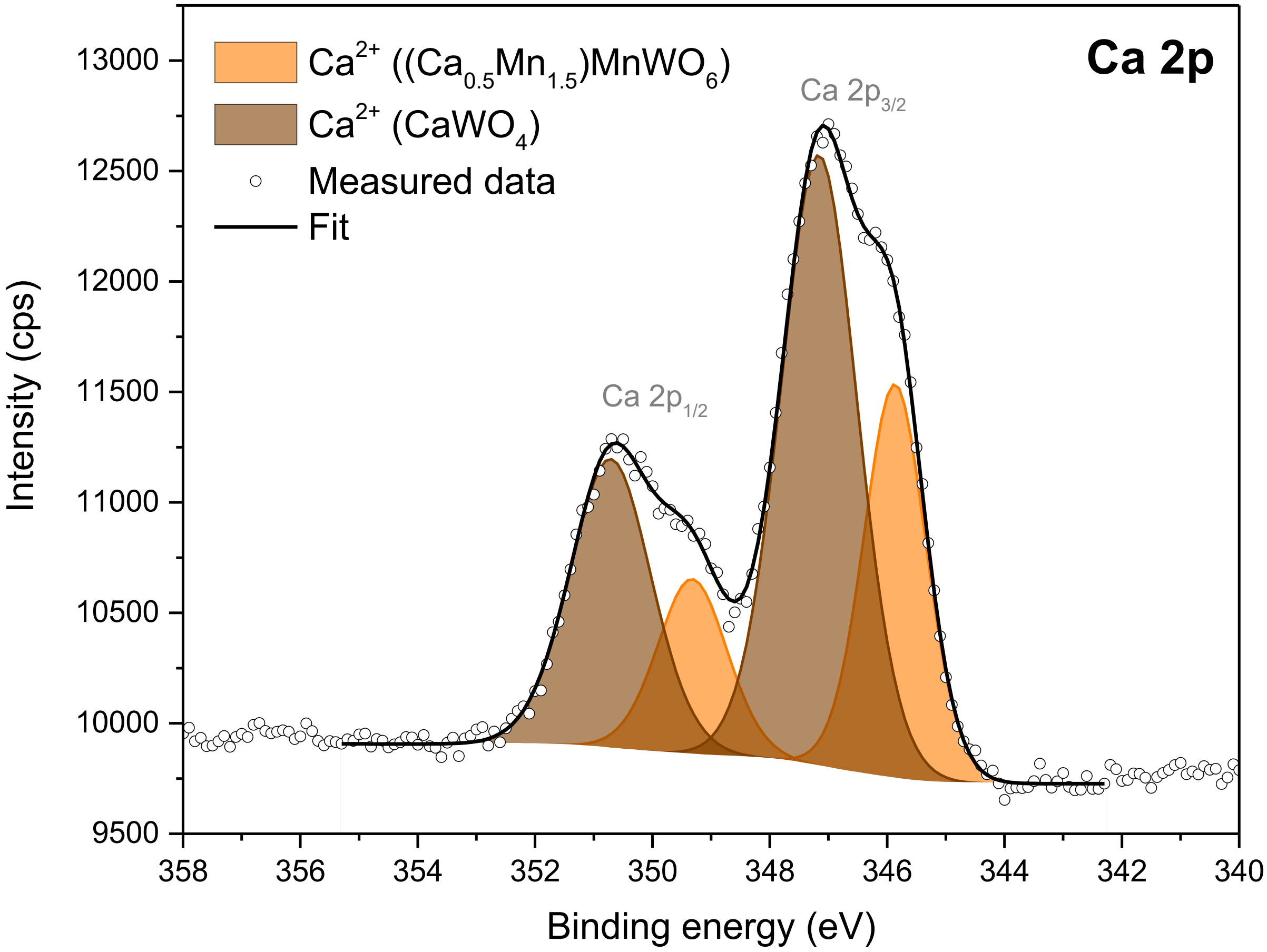

**Figure S10.** The XPS Ca 2p core level region of the powdered $(Ca_{0.5}Mn_{1.5})MnWO_6$. The Ca 2p was fitted by Shirley background and two sets of spin-orbit $2p_{3/2}$ – $2p_{1/2}$ doublets. The brown doublet (347.2 – 350.7 eV) corresponds to $Ca^{2+}$ from $CaWO_4$ [9]. The orange doublet (345.9 – 349.3 eV) has unusually low bidning energy for $Ca^{2+}$ [10]. Similar low-binding-energy components (345.2 – 348.4 eV) have been reported for $Ca^{2+}$ in complex oxide lattice environments in $Ba_2CaWO_6$ [11]. Analogously, we attributed the orange doublet (345.9 – 349.3 eV) to $Ca^{2+}$ from $(Ca_{0.5}Mn_{1.5})MnWO_6$. The spin-orbit splitting and relative intensity ratio in each doublet was fixed to 3.5 eV and 1:2, respectively [10].

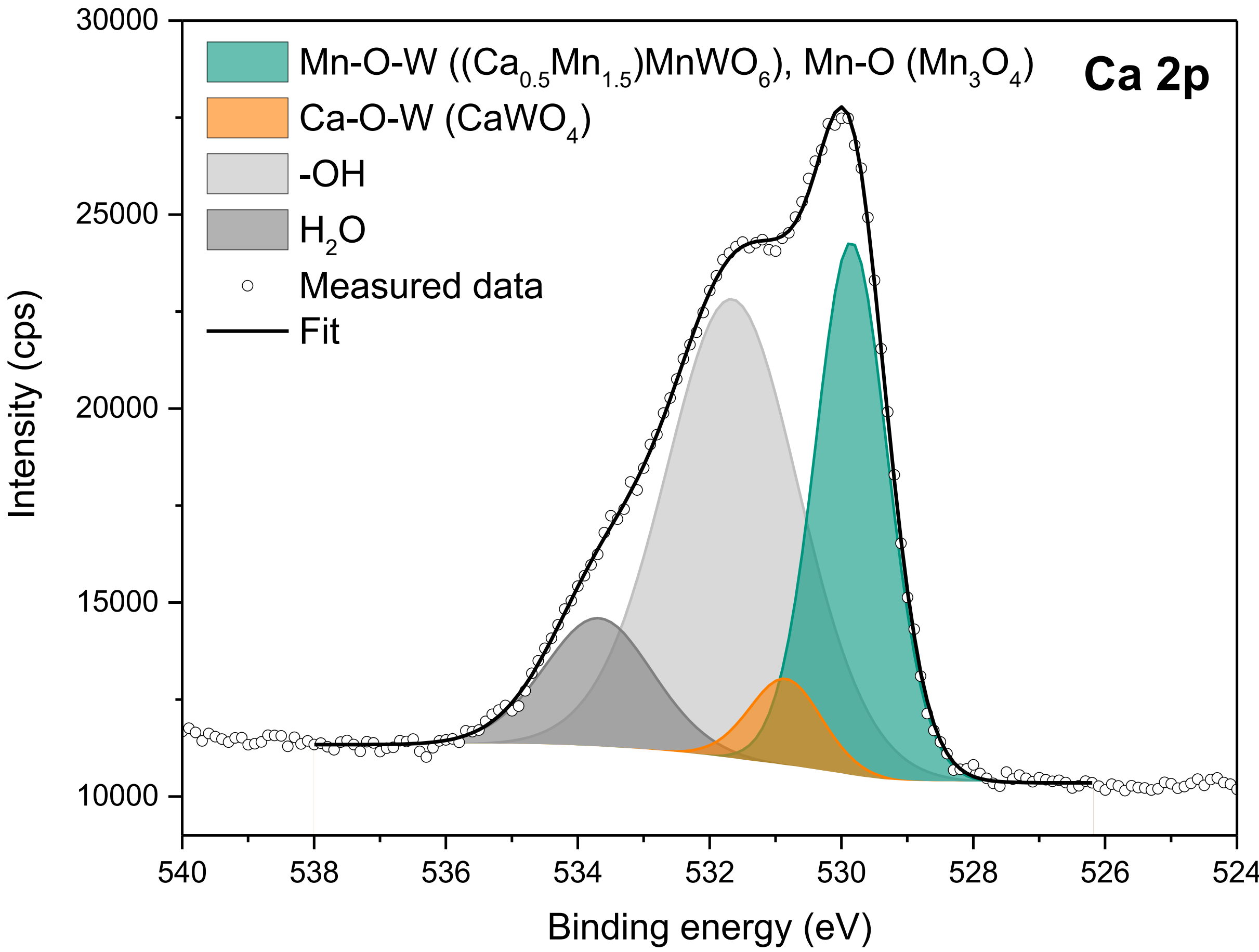

**Figure S11.** The XPS O 1s core level region of the powdered ($(Ca_{0.5}Mn_{1.5})MnWO_6$). The O 1s was fitted by Shirley background and four peaks corresponding to Mn-O-W in $(Ca_{0.5}Mn_{1.5})MnWO_6$ and Mn-O in $Mn_3O_4$ – turqois [4, 7, 8], Ca-O-W in $CaWO_4$– orange [9], adsorbates (hydroxyls, carbonates, etc.) [4, 7, 8] and water [4, 8].

Our conclusions about $Mn^{3+}$ origin are based on a complex XPS analysis. Because interpretation of manganese XPS spectra is not trivial, we first determined the Mn formal valency from Mn 3s core level to be +2.3 (the Mn 3s spectrum is not so heavily affected by the multiple-splitting effect) [6]. Afterwards, we fitted the Mn 2p core level spectrum, which can give more accurate information about the bonding configuration of manganese. The fitting the Mn 2p relies heavily on references because of strong multiple-splitting in the Mn 2p core-level. Because $(Ca_{0.5}Mn_{1.5})MnWO_6$ is a newly studied system, there are no available XPS references. Therefore, we composed our fit from multiplets of known oxides: MnO, $Mn_2O_3$ and $Mn_3O_4$ [4,5]. The following Figure demonstrates our attempts using these references. We can see that the overall shape resembles pure MnO except for a region of the shake-up satellite at 646 eV (1). On the contrary, $Mn_2O_3$ (2) does not fit at all. $Mn_3O_4$ (3) produces relatively good fit, but the resulting formal valency does not agree with the one estimated from Mn 3s. The combination of MnO and $Mn_2O_3$ oxides (4) suppresses $Mn_2O_3$ entirely (the result is the same as in the case of MnO). The best result, including the formal Mn valency (+2.1) can be obtained by combining MnO and $Mn_3O_4$ (5). *Therefore, we concluded that the* $(Ca_{0.5}Mn_{1.5})MnWO_6$ *sample contains mainly* $Mn^{2+}$.

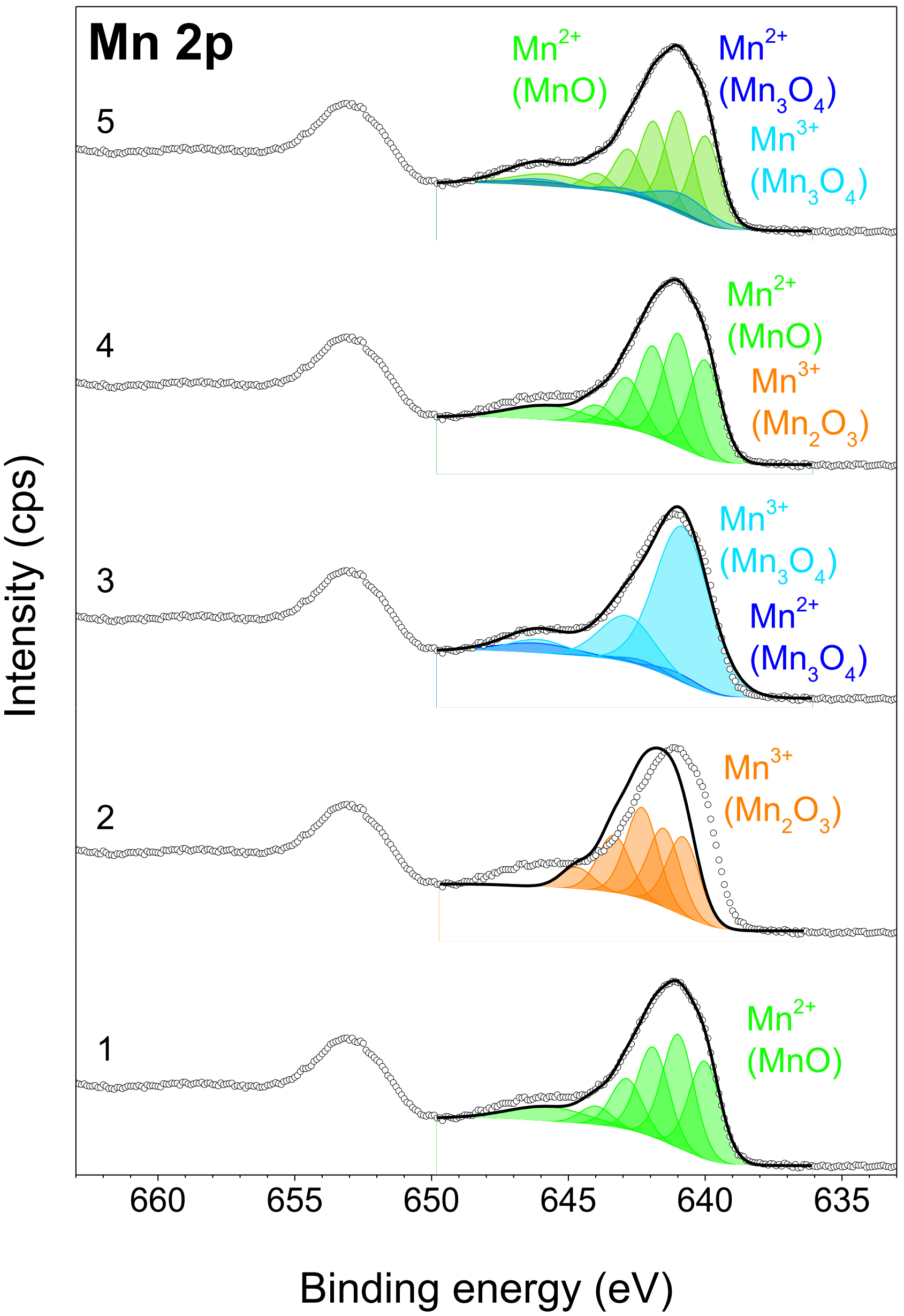

Regarding existence of oxygen vacancies in the system, we believe that the occurrence of $Mn^{2+}$, $Ca^{2+}$ and $W^{6+}$ is not in contradiction with existence of oxygen vacancies in the system. XPS confirmed $Mn^{2+}$ and $W^{6+}$, and Ca was found at unusually low binding energy (between $Ca^{2+}$ and $Ca^0$, which indicates slightly higher effective negative charge compared to $Ca^{2+}$-O). The oxygen vacancies cannot be directly observed by XPS since they do not produce photoelectrons. XPS can only provide indirect evidence of the vacancies, such as OH groups adsorbed on the vacancy site (clearly visible in the O 1s core level spectrum).

When oxygen vacancy is formed in the oxide, two electrons must be accommodated by the system. In easily reducible oxides, such as $TiO_2$, excess electrons are strongly localized on Ti ions, causing electronic reduction ($Ti^{4+} \rightarrow Ti^{3+}$) [12,13]. We did not find any traces of $Mn^0$, $Ca^0$ and $W^5$ in $(Ca_{0.5}Mn_{1.5})MnWO_6$ therefore, the excess electrons left in the system after oxygen vacancy formation should be accommodated elsewhere. There are several possible alternatives:

1) Electrons are trapped at oxygen vacancies sites (F-centers) [13]. This was reported for oxygen-deficient $WO_{3-x}$ where wolfram remains in $W^{6+}$ oxidation state [12]. Additionally, if the oxygen vacancy is formed near $Ca^{2+}$ cations, it could result in increase in its negative effective charge and lowering its binding energy (observed by XPS).

2) Electrons may partially occupy antibonding Ca 3d – O 2p hybridized states which could be a reason for higher conductivity and lower binding energy in Ca 2p [Medvedeva, Shao].

3) Electrons can be delocalized in the W 5d – O 2p hybridized state and increase the conductivity of the sample [12,14].

All above-mentioned explanations are plausible but unfortunately, XPS analysis cannot provide conclusive evidence for any of them. However, we hope that we demonstrated that existence of oxygen vacancies in our system is possible and explains well increased conductivity of the material.

**TABLE SIV.** Fitting parameters of XPS core level regions: binding energy (BE), full-width-in-half-the-maximum (FWHM), relative peak-position-difference (Δ), relative intensity, and relative atomic concentration of compounds within each region (the sum is 100 % for each region).

| Core level | Compound | BE (eV) | FWHM (eV) | Δ (eV) | Relative intensity | Relative concentration (at.%) |
|---|---|---|---|---|---|---|
| Mn $2p_{3/2}$ | **$Mn^{2+}$** $(Ca_{0.5}Mn_{1.5})MnWO_6$ | 634.0 | 1.2 | A [4] | $I_A$ = 1.00 [4] | 78 ± 15 |
| | | 641.0 | 1.2 | A + 1.0 [4] | $I_A$ * 1.16 [4] | |
| | | 641.9 | 1.2 | A + 1.9 [4] | $I_A$ * 0.92 [4] | |
| | | 642.8 | 1.2 | A + 2.8 [4] | $I_A$ * 0.52 [4] | |
| | | 644.0 | 1.2 | A + 4.0 [4] | $I_A$ * 0.20 [4] | |
| | | 645.7 | 3.2 | A + 5.7 [4] | $I_A$ * 0.38 [4] | |
| | **$Mn^{2+}$** $Mn_3O_4$ | 641.0 | 2.1 | B [5] | $I_B$ = 1.00 [5] | 7 ± 2 |
| | | 642.3 | 2.1 | B + 1.3 [5] | $I_B$ * 0.75 [5] | |
| | | 643.3 | 2.1 | B + 2.3 [5] | $I_B$ * 0.40 [5] | |
| | | 644.2 | 2.1 | B + 3.2 [5] | $I_B$ * 0.20 [5] | |
| | | 646.2 | 3.6 | B + 5.2 [5] | $I_B$ * 2.00 [5] | |
| | **$Mn^{3+}$** $Mn_3O_4$ | 640.8 | 2.4 | C [5] | $I_C$ = 1.00 [5] | 15 ± 3 |
| | | 642.8 | 2.4 | C + 2.0 [5] | $I_C$ * 0.30 [5] | |
| | | 646.1 | 2.5 | C + 5.3 [5] | $I_C$ * 2.00 [5] | |
| Mn 3s | **$Mn^{2+}$** | 82.3 | 1.8 | D [6] | $I_D$ = 1.00 [6] | 72 ± 7 |
| | | 88.4 | 2.5 | D + 6.1 [6] | $I_D$ * 0.67 [6] | |
| | **$Mn^{3+}$** | 83.7 | 1.4 | E [6] | $I_E$ = 1.00 [6] | 28 ± 3 |
| | | 89.1 | 2.5 | E + 5.4 [6] | $I_E$ * 0.62 [6] | |
| W 4f | **$W^{6+}$ $4f_{7/2}$** $(Ca_{0.5}Mn_{1.5})MnWO_6$ | 35.1 | 1.0 | F | $I_F$ = 1.00 | 74 ± 7 |
| | **$W^{6+}$ $4f_{5/2}$** $(Ca_{0.5}Mn_{1.5})MnWO_6$ | 37.3 | 1.0 | F + 2.2 [10] | $I_F$ * 0.75 [10] | |
| | **$W^{6+}$ $4f_{7/2}$** $CaWO_4$ | 35.8 | 1.5 | G | $I_G$ = 1.00 | 26 ± 3 |
| | **$W^{6+}$ $4f_{5/2}$** $CaWO_4$ | 38.0 | 1.5 | G + 2.2 [10] | $I_G$ * 0.75 [10] | |
| Ca 2p | **$Ca^{2+}$ $2p_{3/2}$** $(Ca_{0.5}Mn_{1.5})MnWO_6$ | 345.9 | 1.3 | H | $I_H$ = 1.00 | 34 ± 3 |
| | **$Ca^{2+}$ $2p_{1/2}$** $(Ca_{0.5}Mn_{1.5})MnWO_6$ | 349.3 | 1.4 | H + 3.5 [10] | $I_H$ * 0.50 [10] | |
| | **$Ca^{2+}$ $2p_{3/2}$** $CaWO_4$ | 347.2 | 1.6 | J | $I_J$ = 1.00 | 66 ± 6 |
| | **$Ca^{2+}$ $2p_{1/2}$** $CaWO_4$ | 350.7 | 1.6 | J + 3.5 [10] | $I_J$ * 0.50 [10] | |
| O 1s | **Mn-O-W** $(Ca_{0.5}Mn_{1.5})MnWO_6$, **Mn-O** $Mn_3O_4$ | 529.5 | 1.3 | | | 32 ± 3 |
| | **Ca-O-W** $CaWO_4$ | 530.8 | 1.3 | | | 5 ± 1 |
| | **-OH, CO, COOH** | 531.7 | 2.4 | | | 51 ± 5 |
| | **$H_2O$** | 533.7 | 1.9 | | | 12 ± 1 |